\documentclass[a4paper,12pt]{article}

\usepackage{geometry,slashed}
 \usepackage
 {
   amsmath,amssymb,graphicx,tabularx
 }
  \usepackage{soul}
  \usepackage[usenames,dvipsnames]{xcolor}
 
 \definecolor{X575}{rgb}{0.05, 0.7, 0.05}
 
 \usepackage{color}
 \usepackage{hyperref}
 \usepackage{caption,subcaption}
 \usepackage{authblk}
 \usepackage{epstopdf}
 \usepackage{enumerate}
 \usepackage{multirow}
 \usepackage{siunitx}
 \usepackage{float}

  \usepackage{listings}
  \usepackage{calrsfs}
  \DeclareMathAlphabet{\pazocal}{OMS}{zplm}{m}{n}
 
 \newcommand{\nn}{\nonumber}
 \newcommand{\lt}{\lambda_3}
 \newcommand{\lf}{\lambda_4}
 \newcommand{\kt}{\kappa_3}
 
 \newcommand{\kl}{\kappa_\lambda}

\title{Trilinear Higgs coupling determination via single-Higgs differential measurements at the LHC}

\author[1]{Fabio Maltoni\thanks{fabio.maltoni@uclouvain.be}}
\author[2]{Davide Pagani\thanks{davide.pagani@tum.de}}
\author[1]{Ambresh Shivaji\thanks{ambresh.shivaji@uclouvain.be}}
\author[1]{Xiaoran Zhao\thanks{xiaoran.zhao@uclouvain.be}}

\affil[1]{\small Centre for Cosmology, Particle Physics and Phenomenology (CP3), Universit\'{e} Catholique de Louvain, B-1348 Louvain-la-Neuve, Belgium}
\affil[2]{\small Technische Universit\"{a}t M\"{u}nchen, James-Franck-Str. 1, D-85748 Garching, Germany}

\begin{document}
 
 \maketitle
 
  \vspace*{-9cm}
  {
      {{\color{blue}{CP3-17-37}  \hfill
      \color{blue}{MCNET-17-18}
      \hfill \color{blue}{TUM-HEP-1099/17}}
  }
  \vspace*{9cm}

\begin{abstract}
We study one-loop effects induced by an anomalous Higgs trilinear coupling on total and differential rates for the $H\to 4\ell$ decay and some of the main single-Higgs production channels at the LHC, namely,  VBF, $VH$, $t\bar tH$ and $tHj$.  Our results are based on a public code that calculates these effects by simply reweighting samples of Standard-Model-like events for a given production channel. For $VH$ and $t\bar tH$ production, where differential effects are particularly relevant, we include Standard Model electroweak corrections, which have similar sizes but different kinematic dependences. Finally, we study the sensitivity of future LHC runs to determine the trilinear coupling via inclusive and differential measurements, considering also the case where the Higgs couplings to vector bosons and the top quark is affected by new physics. We find that the constraints on the couplings and the relevance of differential distributions critically depend on the expected experimental and theoretical uncertainties. 
\end{abstract}

 \section{Introduction}\label{sec:intro}
Since its discovery in 2012 \cite{Aad:2012tfa, Chatrchyan:2012xdj},  evidence has been steadily accumulating that the properties of the scalar particle at 125 GeV correspond to those of the Higgs boson predicted by the Standard Model (SM) of elementary particles and interactions. ATLAS and CMS experiments have analysed data from several inverse femtobarns of integrated luminosity  at different energies, providing already rather precise measurements of the Higgs couplings to the vector bosons and to the fermions of the third generation~\cite{Khachatryan:2014qaa,Aad:2014lma,Aad:2015iha,Aad:2015gra,Khachatryan:2016vau}. Prospects of the next LHC runs for further  improving the precision on the value of these couplings and for measuring also the couplings to the second-generation fermions  are very good. On the contrary, the situation and prospects for determining the properties of the scalar potential, {\it i.e.}, of the Higgs self-couplings at the LHC are  less clear and therefore the subject of an intense theoretical and experimental activity. 

At low energy, the potential for a scalar particle of mass $m_H$ can be parametrised as a polynomial
\begin{equation}
 V(H) = \frac{1}{2} m_H^2 H^2 + \lt v H^3 + \frac{1}{4}\lf H^4 + {\it O}(H^5), 
\label{Hpotential}
\end{equation}
where $v=(\sqrt{2}G_F)^{-1/2} \sim 246~{\rm GeV}$ is the vacuum expectation value after electroweak-symmetry-breaking  (EWSB). In the SM, renormalisability and gauge invariance dictate that the Higgs potential depends only on two parameters, $\mu$ and  $\lambda$,
 \begin{equation}
  V^{\rm SM}(\Phi) = -\mu^2 (\Phi^\dagger \Phi) + \lambda (\Phi^\dagger \Phi)^2\, , \label{Vsm}
  \end{equation}
  where $\Phi$ is the Higgs doublet.
After EWSB the potential in eq.~\eqref{Vsm} gives rise to the mass of the physical Higgs boson $m_H$, which, together with the vacuum expectation value $v$, are related to $\mu$ and $\lambda$ via  
$\mu^2=m_H^2/2$ and $\lambda=m_H^2/(2v^2)$. As a result, by fixing $\mu$ and $\lambda$, in  the SM the Higgs self couplings are completely determined, leading in eq.~\eqref{Hpotential} to $\lambda_i=\lambda_i^{\rm SM}$ with $\lt^{\rm SM} = \lf^{\rm SM} = \lambda$ and $\lambda_i^{\rm SM} = 0$ for $i\ge 5$, at LO. Thus,  with $m_H=125~{\rm GeV}$,
\begin{equation}
\lt^{\rm SM} = \lf^{\rm SM} \simeq 0.13\, .  
\end{equation}
On the other hand, new physics could modify the Higgs potential at low energy, by altering the value of $\lt$ (or in general $\lambda_i$ for the $i$-point Higgs self couplings) without affecting the value of $m_H$ and $v$. This can be realised either directly ({\it e.g.} by extending the scalar sector) or indirectly (due to the exchange of new  virtual states). In addition, modifications in the self interactions would be induced if the Higgs boson is a composite state.

Since double Higgs production directly depends on the Higgs trilinear coupling at LO, it is the standard process for studying $\lt$ at the LHC. However, the cross section of its main production channel,  the gluon fusion, is only about 35 fb at 13 TeV~\cite{Maltoni:2014eza,deFlorian:2015moa,Borowka:2016ehy}, so it is much smaller than single Higgs production cross section, which is about 50 pb~\cite{Anastasiou:2016cez}. Several phenomenological studies have been performed on the determination of $\lt$ via the relevant experimental signatures emerging from 
this process: $b\bar{b} \gamma \gamma$~\cite{Baur:2003gp, Baglio:2012np,Yao:2013ika, Barger:2013jfa, Azatov:2015oxa, Lu:2015jza}, $b\bar{b}\tau \tau$~\cite{Dolan:2012rv, Baglio:2012np}, 
$b\bar{b}W^+W^-$~\cite{Papaefstathiou:2012qe} and
$b\bar{b}b\bar{b}$~\cite{deLima:2014dta, Wardrope:2014kya,Behr:2015oqq}. Also $t{\bar t}HH$~\cite{Englert:2014uqa,Liu:2014rva} and 
$VHH$~\cite{Cao:2015oxx} production processes have beeen considered.
Nevertheless, given the complexity of a realistic experimental set-up, the final precision that could be achieved on the determination of $\lt$ is still
unclear.  
On the contrary, it is established that at the LHC perspectives of inferring information on $\lf$ from the triple Higgs production are quite bleak 
\cite{Plehn:2005nk, Binoth:2006ym}, due to the smallness of the
corresponding cross section~\cite{Maltoni:2014eza,deFlorian:2016sit} together with a rather weak dependence on this parameter. 
Even at a future 100 TeV proton--proton collider a considerable amount of integrated luminosity will be necessary in order to obtain rather loose bounds~\cite{Chen:2015gva,Kilian:2017nio,Fuks:2017zkg}.

At the moment the strongest experimental bounds on non-resonant double-Higgs production have been obtained in the CMS analysis of the $b\bar{b}\gamma \gamma$ signature \cite{CMS:2017ihs}, where cross-sections larger than about 19 times the predicted SM value have been excluded.  However, exclusion limits on $\lt$ are found to be strongly dependent on the value of the top-Higgs coupling and they are of the order $ \lt <-9 ~\lt^{\rm SM}$ and $\lt > 15 ~\lt^{\rm SM}$ for the SM-like case. These new limits, together with the slightly weaker ones from the ATLAS $b\bar{b}b\bar{b}$ \cite{ATLAS:2016ixk} and CMS $b\bar{b}\tau \tau$ \cite{CMS:2017orf} analyses at 13 TeV, improve the results from 8 TeV data ($\lt< -17.5 ~\lt^{\rm SM}$ and $\lt > 22.5 ~ \lt^{\rm SM}$ \cite{Khachatryan:2016sey}), however, also with a high integrated luminosity (HL) of 3000 fb$^{-1}$, a further improvement may not be so tremendous. The most optimistic {\it experimental} studies for HL-LHC suggest that it could be possible to exclude values in the range $\lt<-1.3 ~\lt^{\rm SM}$ and
$\lt>8.7 ~ \lt^{\rm SM}$ via the $b\bar{b} \gamma \gamma$
signatures \cite{ATL-PHYS-PUB-2014-019}. Additional and complementary strategies for the determination of $\lt$ are thus desirable at the moment.

In ref.~\cite{McCullough:2013rea} an indirect method of measuring $\lt$ via EW radiative corrections in $e^+e^- \to ZH$ process was proposed. 
Recently,  the same idea has also been extended to the LHC, by (globally) studying $\lt$-dependent EW corrections in single Higgs production and decay processes~\cite{Gorbahn:2016uoy,Degrassi:2016wml,Bizon:2016wgr,DiVita:2017eyz}. Moreover, limits on $\lt$ can also be derived by two-loop effects in EW precision observables~\cite{Degrassi:2017ucl,Kribs:2017znd} such as the measurements   of $m_W$  and of the $S, T$ oblique parameters. These studies have confirmed that indirect bounds on $\lt$ can be competitive with the direct ones inferred from the di-Higgs production channel. For example, a simple one-parameter fit to the signal strengths measurements at 8 TeV~\cite{Khachatryan:2016vau} gives $-9.4~\lt^{\rm SM}<~\lt<17.0~\lt^{\rm SM}$~\cite{Degrassi:2016wml}, comparable to the current best constraints from the $b\bar{b}\gamma\gamma$  CMS measurement mentioned above. In both analyses no other deviations for the Higgs couplings are considered. The usefulness of a joint analysis of $\lt$ indirect effects on single-Higgs production and direct effects on di-Higgs production has been discussed and quantified in ref.~\cite{DiVita:2017eyz}. As already suggested in ref.~\cite{Degrassi:2016wml,Bizon:2016wgr}, the role of differential distributions and their non-flat dependence on $\lt$ has been proved to be crucial in ref.~\cite{DiVita:2017eyz}, especially when not only anomalous $\lt$ effects  but also modifications of the couplings to the other particles are considered, as expected in a general Effective-Field-Theory (EFT) approach.  

\medskip

The purpose of this work is threefold. First, we present an automated public code for generating events including $\lt$ effects at one loop, thus allowing the study of differential effects in VBF, $VH$ and $t \bar t H$ production\footnote{The impact of differential distributions in gluon-fusion is not  studied here as one would need to consider the effects of the trilinear coupling in $H+1$ jet. This  involves the calculation of $2\to 2$ two-loop amplitude with four independent scales, which is not yet feasible.} and all the relevant Higgs decays. The code is based on two independent and procedurally different implementations in the {\sc MadGraph5\_aMC@NLO} framework~\cite{Alwall:2014hca}.~\footnote{The code can be found at the webpage:\\
 \url{ https://cp3.irmp.ucl.ac.be/projects/madgraph/wiki/HiggsSelfCoupling}.
}
 Second, with the help of this code, we extend at the differential level the results of ref.~\cite{Degrassi:2016wml}, where all the relevant single Higgs production ($gg{\rm F}$, ${\rm VBF}$, $VH$, $t\bar tH$) and decay  channels ($\gamma\gamma$, $VV^*$, $ff$, $gg$) have been analysed and included in a global fit only at the inclusive level. Indeed, in ref.~\cite{Degrassi:2016wml} the usefulness of differential distributions has already been explored but only for the case of $VH$ and $t \bar t H$ production, providing the results on which the analysis in ref.~\cite{DiVita:2017eyz} rely. Differential information for  VBF and $VH$ production has been presented also in ref.~\cite{Bizon:2016wgr}. Here we scrutinise all the relevant distributions that are potentially affected by anomalous $\lt$ effects, presenting for the first time detailed results at the differential level for $t \bar t H$ production, for the $H\to 4\ell$ decay and also for the $tHj$ process, for which also inclusive results are new. Moreover, for the case of $VH$ and $t \bar t H$ production, where loop-induced $\lt$ effects are not flat, we repeat the calculation of NLO EW corrections \cite{Ciccolini:2003jy, Denner:2011id,Granata:2017iod,Frixione:2014qaa,Frixione:2015zaa,Yu:2014cka} in the SM, which are  also not flat, in order to check the robustness of our strategy. As expected, we find that NLO EW corrections are essential for a precise determination of anomalous $\lt$ effects, but also that they do not jeopardise the sensitivity of indirect $\lt$ determination. We use for the calculation the EW extension  of the automated  {\sc MadGraph5\_aMC@NLO} framework that has already been used and validated in refs.~\cite{Frixione:2014qaa,Frixione:2015zaa, Badger:2016bpw, Pagani:2016caq, Frederix:2016ost, Czakon:2017wor}.~\footnote{To our knowledge, NLO EW corrections to the $tHj$ process are calculated for the first time here.}

  Finally we perform a fit for $\lt$ based on the future projections of ATLAS-HL for single-Higgs production and decay at 14 TeV~\cite{ATL-PHYS-PUB-2013-014, ATL-PHYS-PUB-2014-012}. We consider the effects induced on the fit by additional degrees of freedom, namely anomalous Higgs couplings with the vector bosons and/or the top quark. We investigate the impact on the fit of three different factors: the differential information, the experimental and theoretical uncertainties,  and the inclusion of the  two aforementioned   additional degrees of freedom. We find that in a global fit, including all the possible production and decay channels, two additional degrees of freedom such as those considered here do not preclude the possibility of setting sensible $\lt$ bounds, especially, they have a tiny impact on the upper bound for positive $\lt$ values. On the contrary the role of differential information may be relevant, depending on the assumptions on  experimental and theoretical uncertainties.

The structure of the paper is the following.
In sec.~\ref{sec:c1-calc} we briefly repeat and comment the main formulas that are relevant in the study of one-loop induced $\lt$ effects. Differential results for the various processes that have been mentioned before  are given in sec.~\ref{sec:c1-dist}. In sec.~\ref{sec:nlo-ew} we present the extension of the analysis of $\lt$ effects including NLO EW corrections and we study the impact on differential distributions, as well as inclusive rates. In sec.~\ref{sec:global-fit} we introduce the framework for the fit and discuss the results obtained. Details about the statistical treatment of uncertainties in the fit are reposted in Appendix \ref{appic}.

 \section{Technical setup}
 \label{sec:c1-calc}
\subsection{Self couplings effects in single Higgs production and decays at one loop} 
 Higgs self-couplings can be studied in a model-dependent approach, {\it e.g.}, choosing a particular UV-complete scenario, or in  a model-independent approach, as done in this work. However, there are different ways in which the modifications of trilinear and quartic couplings can be parametrised and they rely on different theoretical assumptions.
 If new physics is at scales sufficiently higher than the energies where measurements are performed, the SMEFT offers a consistent and model-independent way of organising generic deformations to the Higgs interactions.  Moreover, radiative corrections can be consistently performed within this framework. However, it has been shown~\cite{Degrassi:2016wml} that in the case of the trilinear coupling and at the order  we are considering, one-loop corrections for single Higgs processes, adding higher dimensional operators that only affect the Higgs self couplings (the  $(\Phi \Phi ^\dag)^n$ operators with $n>2$)  or directly introducing an anomalous coupling
\begin{equation}
 \lt = \kt \lt^{\rm SM} \label{eq:np}\,,
\end{equation}
are two fully equivalent approaches for deforming the SM Higgs potential. In other words, regarding the Higgs self-couplings, differences between an EFT and an anomalous coupling parametrisation will arise only for final states featuring more Higgs bosons and/or at higher loop -level, {\it i.e.}, with the appearance of higher-point interactions (starting from the quadrilinear). 

It is essential to note that the single Higgs production and the decay channels are not sensitive to $\lf$ at one loop. For this reason, although results are delivered in terms of $\lt$, they can be easily translated in terms of the Wilson coefficient in front of the dimension-6 operator $(\Phi \Phi ^\dag)^3$, or those for higher-dimension $(\Phi \Phi ^\dag)^n$ operators. In the following we briefly repeat and comment the main formulas that have been introduced and discussed at length in ref.~\cite{Degrassi:2016wml}; very minor modifications are present in the notation and in the definition of the corresponding quantities.   
  \begin{figure}[t]
  \begin{center}
 \includegraphics[scale=0.5]{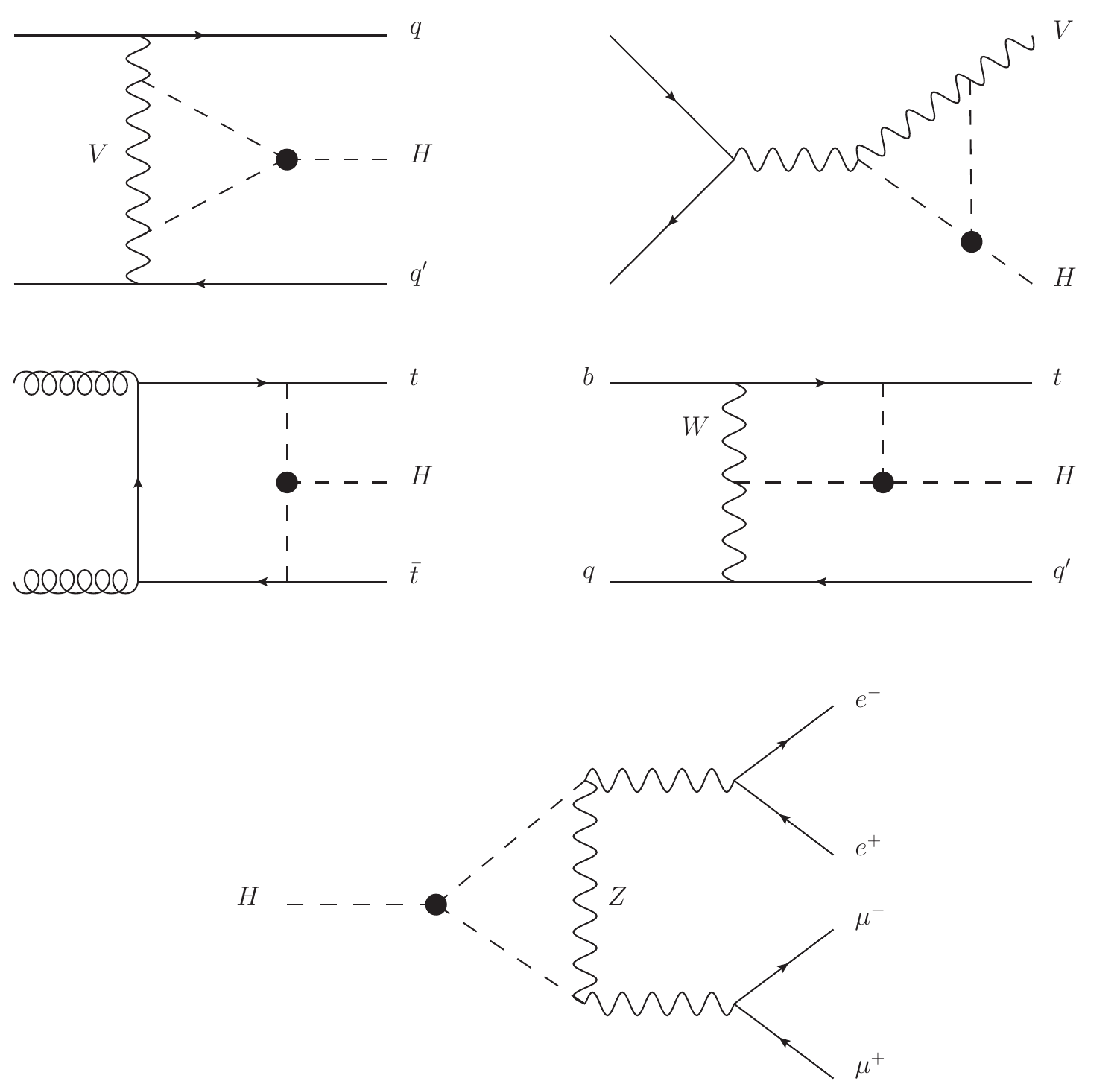}
   \caption{ Representative one-loop diagrams in single Higgs processes with anomalous trilinear coupling. Differential information on $gg$F requires the calculation of EW two-loop amplitudes for $Hj$ production, which is not yet feasible with the current technology. }
   \label{fig:trilinear-ol}
   \end{center}
 \end{figure}
 
The $\lt$-dependent part of the NLO EW corrections to single Higgs processes are gauge invariant and  finite. 
These contributions can be organised in two categories: 
a universal part proportional to $(\lt)^2$, which arises from the wave-function renormalisation of external Higgs boson and thus does not depend on the kinematics, and 
a process-dependent part linear in $\lt$, which is also sensitive to the kinematics. 
 In presence of modified 
trilinear coupling, 
the master formula for the $\lt$-dependence of a generic observable $\Sigma$ 
(total/differential cross section or decay width) can be written as 
\begin{eqnarray}
 \Sigma_{\lt}^{\rm BSM} &=& Z_H^{\rm BSM} \Sigma_{\rm LO} (1 + \kt C_1 + \delta Z_H) \,, \label{master}
\end{eqnarray}
where  $C_1$ is the process- and kinematic-dependent component and $\Sigma_{\rm LO}$ stands for the LO prediction including any factorisable higher-order correction.  In particular, we assume that QCD corrections do in general factorise, which, for the $VH$ and VBF case,  has been shown to be a correct approach up to NNLO in ref.~\cite{Bizon:2016wgr}.\footnote{As the weak loops considered here are always characterised by scales of the order of the mass of the heavy particles in the propagators (weak bosons, top quarks and the Higgs) while QCD corrections at threshold are typically dominated by lower scales, factorisation is a reasonable working assumption.} Representative diagrams contributing to the $C_1$ for the different processes are depicted in Fig.~\ref{fig:trilinear-ol}.

In eq.~\eqref{master},  at variance with the case of $\Sigma_{\lt}^{\rm NLO}$ in ref.~\cite{Degrassi:2016wml},  
 the universal component $Z_H^{\rm BSM}$ corresponds to the wave function renormalisation where we have resummed {\it only} the new-physics contributions at one loop, 
 \begin{eqnarray}
 Z_H^{\rm BSM} &=& \frac{1}{1-(\kt^2-1) \delta Z_H}\,, \label{ZHBSM} \\
 \delta Z_H &=& -\frac{9}{16\sqrt{2}\pi^2} \left(\frac{2\pi}{3\sqrt{3}}-1\right) G_\mu m_H^2  = -1.536\times 10^{-3}.
\end{eqnarray}
The SM component is directly included at fixed NLO via the  $\delta Z_H$ term appearing in eq.~\eqref{master}.
 Numerically, the difference between eq.~\eqref{master} and $\Sigma_{\lt}^{\rm NLO}$ in ref.~\cite{Degrassi:2016wml} is at sub-permill level and thus negligible. On the other hand, in the limit  $\kt\to1$, $Z_H^{\rm BSM}\to1$ and thus   $\Sigma_{\lt}^{\rm BSM}$ goes to the SM case at fixed NLO
\begin{eqnarray}
 \Sigma_{\lt}^{\rm SM} &=& \Sigma_{\rm LO} (1 + C_1 + \delta Z_H)\,.
\end{eqnarray}
This is particularly convenient for the discussion in sec.~\ref{sec:nlo-ew}, where we will analyse NLO EW corrections in the SM in conjunction with $\lt$-induced effects. 
 In conclusion, the relative corrections due to trilinear coupling can be expressed as
\begin{eqnarray}
 \delta\Sigma_{\kt} = \frac{\Sigma_{\lt}^{\rm BSM}-\Sigma_{\lt}^{\rm SM}}{\Sigma_{\rm LO}} 
                          = (Z_H^{\rm BSM}  -1) (1+\delta Z_H) + (Z_H^{\rm BSM}  \kt -1)C_1,  \label{dsigmakt}
\end{eqnarray}
which manifestly goes to zero in the $\kt\to1$ limit. 

Numerical values of $C_1$ at the inclusive level for the processes considered in this work are reported in Tab.~\ref{tab:c1-xs}. The calculation of $C_1$ for single-top-Higgs production, which appears for the first time here, is non-trivial and discussed in sec.~\ref{sth}. The range of validity of eq.~\eqref{dsigmakt} has been identified in ref.~\cite{Degrassi:2016wml} as $|\kt| < 20$, given the values of $ \delta Z_H$ and $C_1$ in  Tab.~\ref{tab:c1-xs}. As we will see, at the differential level this limit may be too loose since $C_1$ can receive large enhancements (see sec.~\ref{ttHsec}). On the other hand, we believe that the constraint $|\kt| \lesssim 6$ identified in ref.~\cite{DiLuzio:2017tfn} is appropriate for inclusive double Higgs production, but is too strong for the case of single-Higgs production. Indeed the violation of perturbativity for the $HHH$ vertex is kinematic dependent and the condition $|\kt| \lesssim 6$ arises from the configuration with two $H$ bosons on-shell and the third one with virtuality slightly larger than $2m_{H}$. This is the kinematic configuration present above the threshold in double Higgs production, where the bulk of its cross section comes from, but is never present in   single Higgs production, since only one Higgs boson can be on-shell in the $HHH$ vertex appearing at one loop.
 \begin{table}
\begin{center}
\begin{tabular}{|c |c| c| c| c| c| c|| c|}
\hline
 Channels  &$gg$F  & VBF & $ZH$ & $WH$ &  $t\bar tH$ &  $tHj$ & $H\to4\ell$ \\ 
\hline
 $C_1(\%)$ &0.66  &0.63 &1.19 &1.03  &3.52 &0.91 & 0.82\\
\hline
\end{tabular}
\caption{ $C_1$ for different Higgs production processes at 13 TeV LHC and the $H\to 4\ell$ decay.}
\label{tab:c1-xs}
\end{center}
\end{table}
\subsection{Automated codes for the event-by-event calculation of $C_1$}
While $\delta Z_H$ is a universal quantity,  $C_1$ is process and kinematics dependent. 
We have employed two independent methods for the computation of $C_1$ for  differential cross sections and decay rates. They  correspond to the two different codes publicly available, which agree within their numerical accuracy. In the first we parametrise the finite one-loop corrections due to the trilinear Higgs self coupling 
 as form factors that are function of the external momenta. One-loop integrals are computed using the  {\tt LoopTools} package~\cite{Hahn:1998yk} and
 the form factors are implemented as effective new vertices in 
a dedicated UFO model file~\cite{Degrande:2011ua}, which is then used in {\sc MadGraph5\_aMC@NLO}~\cite{Alwall:2014hca}. As a result,  parton level events can be generated including ${\mathcal O}(\lt)$ effects and any interesting observable analysed. 
Our current implementation of form factors allows the computation of differential $C_1$ for VBF, $VH$  and $H \to 4l$ processes. At the order of accuracy of our calculation, $gg$F production and all the other $1 \rightarrow 2$ decays do not have a kinematic dependence for $C_1$; results at the inclusive level are sufficient for any kinematic configuration and thus taken from ref.~\cite{Degrassi:2016wml}.

On the other hand, the implementation of form factors for $t \bar t H$ and $tHj$ processes would be quite cumbersome as there 
 are many one-loop integrals that contribute. A different strategy, based on reweighting, has therefore been devised. With this second method, one starts by generating a sample of (unweighted) parton-level events at leading order. These events are then used as input for a code  that computes the momentum-dependent weight 
  \begin{equation}\label{eq:rw}
 w_i = \frac{2 \Re ({\cal M}^{0 *} {\cal M}^{1}_{\lambda^{\rm SM}_3} )}{|{\cal M}^{0}|^2}\,,
 \end{equation}
 where, following the notation of  ref.~\cite{Degrassi:2016wml}, ${\cal M}^{0 }$ refers to the tree-level amplitude and ${\cal M}^{1}_{\lambda^{\rm SM}_3}$ to the SM virtual corrections depending on $\lt$.  Then, LO events are reweighted by multiplying the weight of each event $i$ by the corresponding $w_i$. 
In this way, also with this method it is possible to calculate $C_1$ for any differential distribution.
 
  The required one-loop matrix elements are computed using the capabilities of {\sc MadGraph5\_\-aMC@NLO} for evaluating loop diagrams~\cite{Hirschi:2011pa,Frederix:2009yq,Frixione:2014qaa,Frederix:2016ost}. For each process, we use diagram filters in order to select the relevant one-loop matrix elements featuring the Higgs self coupling. 
 We find this method much faster and efficient than the one based on form-factors also 
 for $VH$ and VBF processes, thus we actually employ it for deriving all the results presented in  this work and we suggest the usage of this version of the code.
 Note, however, that the other method offers at least in principle the possibility of explicitly including NLO QCD corrections on top of $\lt$-induced effects for VBF and $VH$ production.
 \section{Results for differential distributions}
 \label{sec:c1-dist}
 \begin{figure}[t]
\begin{subfigure}{0.5\linewidth}
  \includegraphics[width=7.8cm,clip]{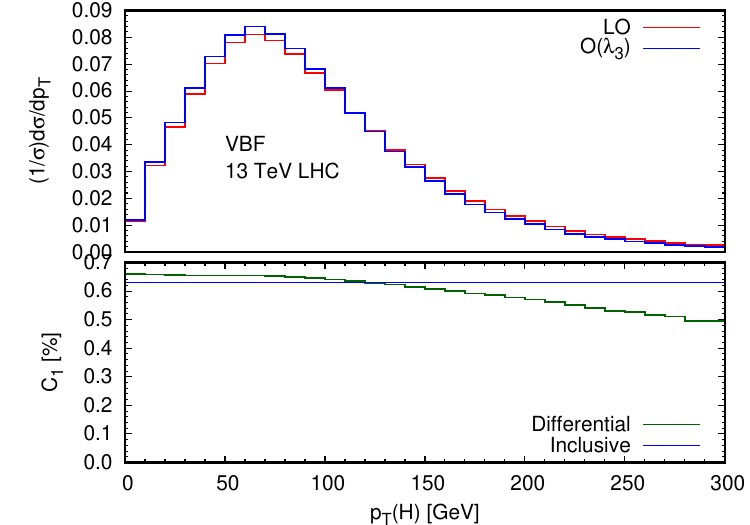}
\end{subfigure}
\begin{subfigure}{0.5\linewidth}
  \includegraphics[width=7.8cm,clip]{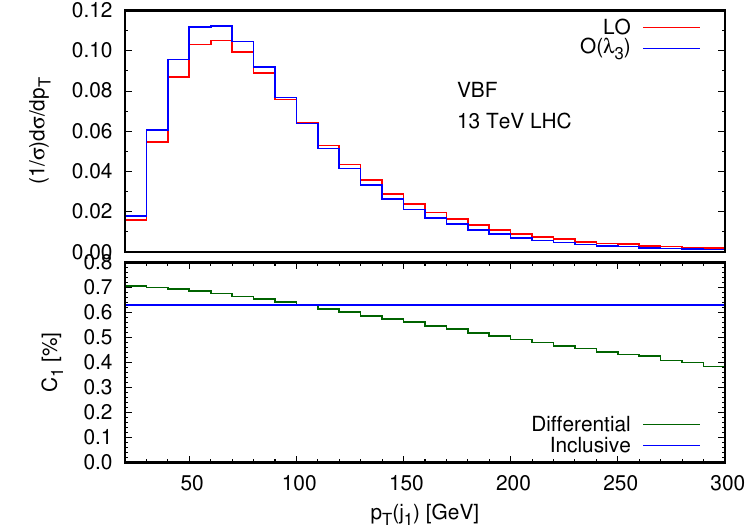}
\end{subfigure} 
 \begin{subfigure}{0.5\linewidth}
   \includegraphics[width=7.8cm,clip]{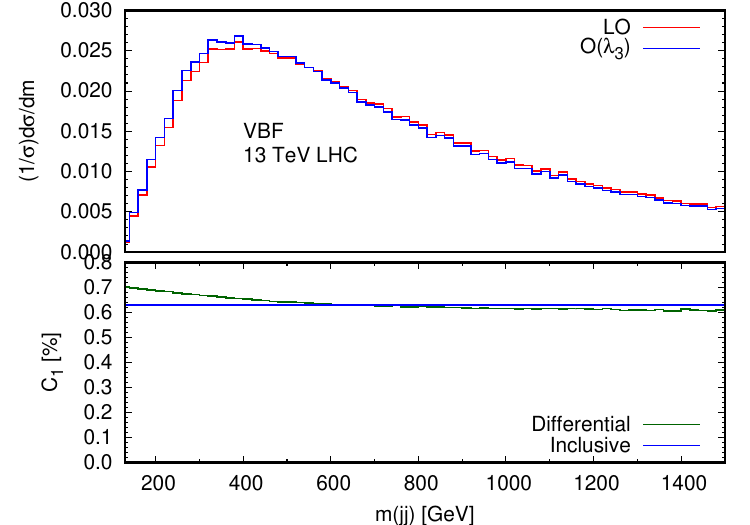}
 \end{subfigure}
\begin{subfigure}{0.5\linewidth}
  \includegraphics[width=7.8cm,clip]{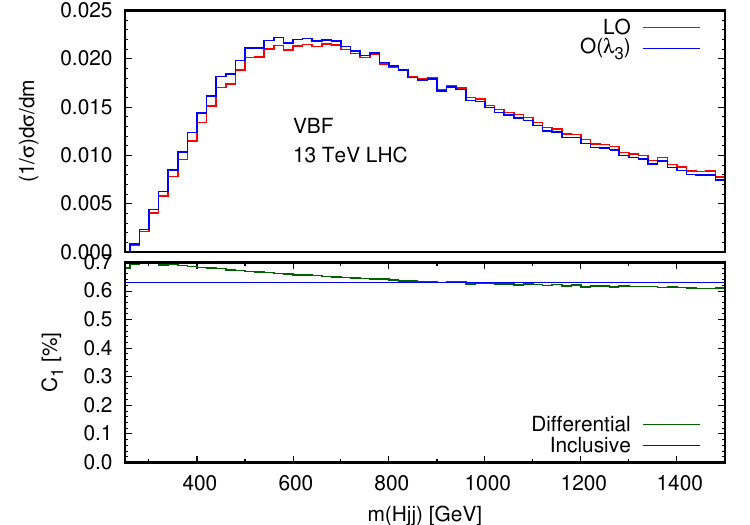}
\end{subfigure} 
\caption{ Effect of ${\mathcal O}(\lt)$ correction in VBF at 13 TeV LHC. Upper panel: 
         normalised distributions at LO (red) and at ${\mathcal O}(\lt)$ (blue). 
          Lower panel: $C_1$ at the differential (green) and inclusive (blue) level.} 
\label{fig:vbf-lo-l3}       
\end{figure}
 Results are presented in this section and have been obtained with the following input parameters,  
  \begin{eqnarray}
G_\mu &=& 1.1663787\times  10^{-5}~{\rm GeV}^{-2},~ m_W = 80.385~{\rm GeV},~m_Z = 91.1876~{\rm GeV}\nn \\
  m_H &=& 125~{\rm GeV}, ~m_t = 172.5~{\rm GeV}\, ,
  \end{eqnarray}
    which are taken from ref.~\cite{deFlorian:2016spz}.
 We use as PDF set the {\tt PDF4LHC2015} distributions with the factorisation scale at   $ \mu_F =  \frac{1}{2} \sum_i m (i)$, where  $m(i)$ are the masses  of the particles $i$ in the final state.~\footnote{As discussed in ref.~\cite{Degrassi:2016wml}, the choice of the factorisation scale has a negligible effects on $C_1$ at inclusive level. The effect is even smaller at differential level.}
 In the following subsections we provide  differential results for various relevant observables in VBF, $VH$, $t \bar t H$ and $tHj$ production channels and in the $H \to 4l$ 
 decay channel. Each plot has the layout that is described in the following. The upper panel displays the 
 LO distribution (red) and ${\mathcal O}(\lt)$ corrections alone (blue), both {\it normalised} by their value for the total cross section. In other words, we compare the shape of LO distributions with the shape of the contributions induced by $C_1$ in eq.~\eqref{master}, which is thus independent on the value of $\kappa_3$. The lower panel display $C_1$ both at differential level (green) and for the  total cross section/decay (blue). The latter values are also summarised in Tab.~\ref{tab:c1-xs} and will be used in the sec.~\ref{sec:global-fit} for the representative fit results.
 \subsection {VBF}
 \begin{figure}[t]
  \begin{center}
\begin{subfigure}{0.49\linewidth}
  \includegraphics[width=7.8cm,clip]{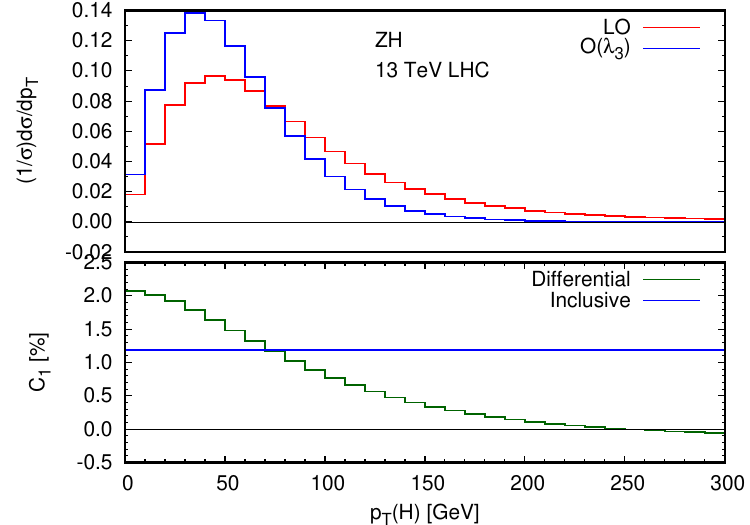}
\end{subfigure}
\begin{subfigure}{0.49\linewidth}
  \includegraphics[width=7.8cm,clip]{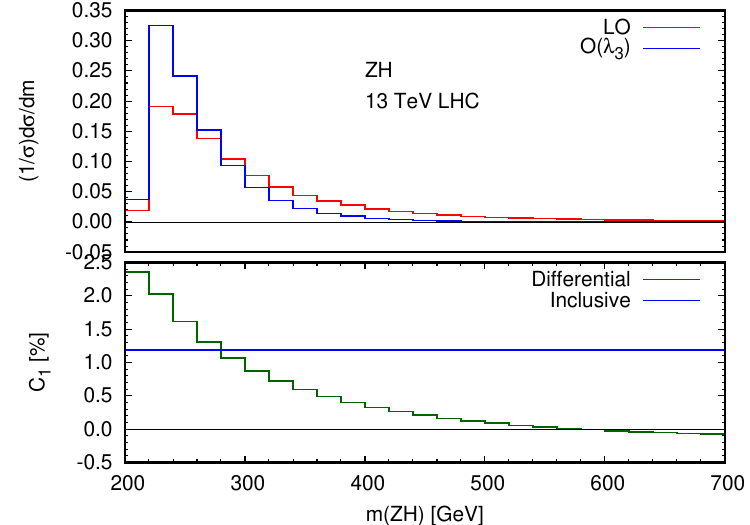}
\end{subfigure} 
\begin{subfigure}{0.49\linewidth}
  \includegraphics[width=7.8cm,clip]{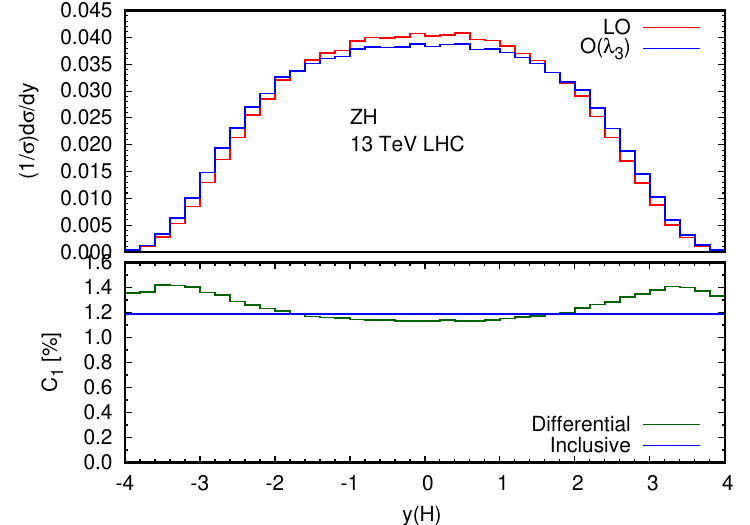}
\end{subfigure}  
\begin{subfigure}{0.49\linewidth}
  \includegraphics[width=7.8cm,clip]{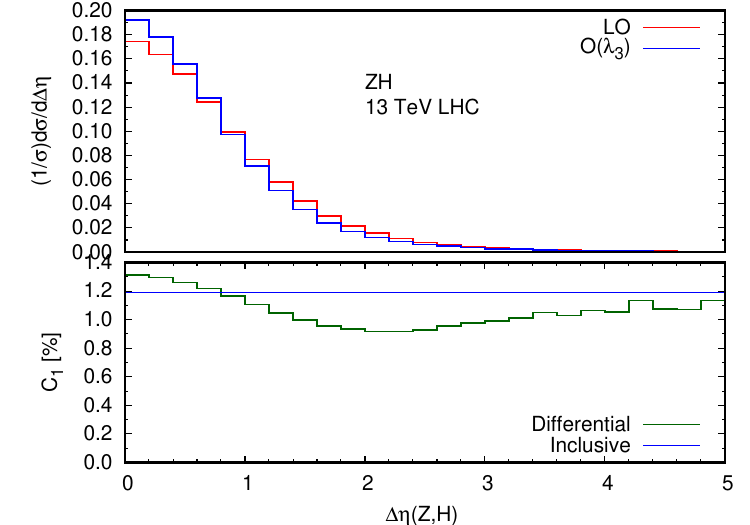}
\end{subfigure}  
\caption{ Effect of ${\mathcal O}(\lt)$ correction in $ZH$ at 13 TeV LHC. Upper panel: 
         normalised distributions at LO (red) and at ${\mathcal O}(\lt)$ (blue). 
          Lower panel: $C_1$ at the differential (green) and inclusive (blue) level.} 
\label{fig:zh-lo-l3}       
\end{center}
\end{figure}
 Vector boson fusion is generated by requiring EW production of Higgs plus two jets, which includes also $VH$ configurations with the vector boson $V$ decaying into two jets. We effectively eliminate $VH$ contributions by applying the following kinematic cuts \cite{deFlorian:2016spz} on the two final-state jets,
 \begin{eqnarray}
  p_T^j > 20~{\rm GeV},~|y_j| < 5,~ |y_{j_i}-y_{j_2}| > 3,~ M_{jj} > 130~{\rm GeV}.
 \end{eqnarray}
In Fig.~\ref{fig:vbf-lo-l3}, we present $C_1$ for representative distributions, namely,  $p_T(H)$, $p_T(j_1)$, $m(jj)$ and $m(Hjj)$. In fact, we have checked that similar effects characterise other observables, which however we do not show.  As already noticed in refs.~\cite{Degrassi:2016wml, Bizon:2016wgr} the value of $C_1$ is not particularly large and rather flat for all the distributions shown here;  $C_1=0.63 \%$ for the total cross section and never exceeds  $0.70\%$ at the differential level. At variance with the case of $VH$ and $t \bar t H$ considered in the following, loop corrections featuring trilinear Higgs self couplings involve Higgs propagators connecting the final-state Higgs and internal $V$ propagators. Thus, no Sommerferld enhancement  is present at threshold.
In this respect, the interest of VBF for what concerns the indirect determination of $\lt$ is mostly limited to the shift in the total rate,  which, even though modest, is anyway relevant. Indeed, VBF is the channel with the second largest cross section and the smallest of the theory uncertainties  \cite{deFlorian:2016spz}, as can also be seen in Tab.~\ref{tab:th-err} in Appendix \ref{appic}. 
 \subsection{${\bf VH}$}
%
\begin{figure}[t]
\begin{subfigure}{0.49\linewidth}
  \includegraphics[width=7.8cm,clip]{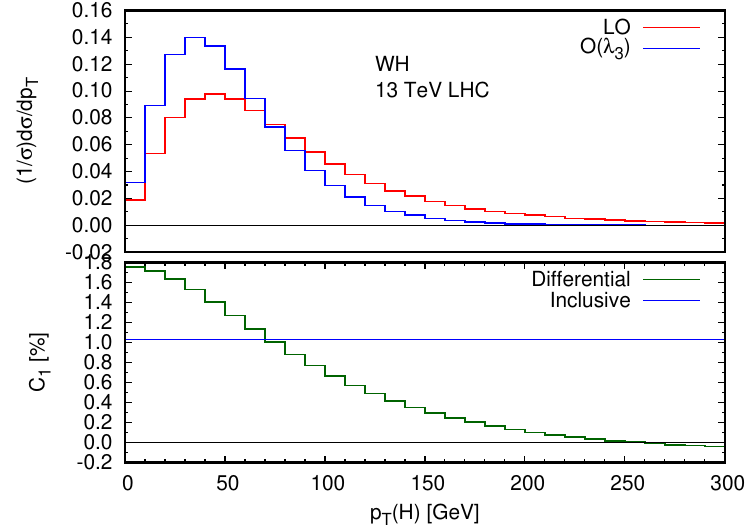}
\end{subfigure}
\begin{subfigure}{0.49\linewidth}
  \includegraphics[width=7.8cm,clip]{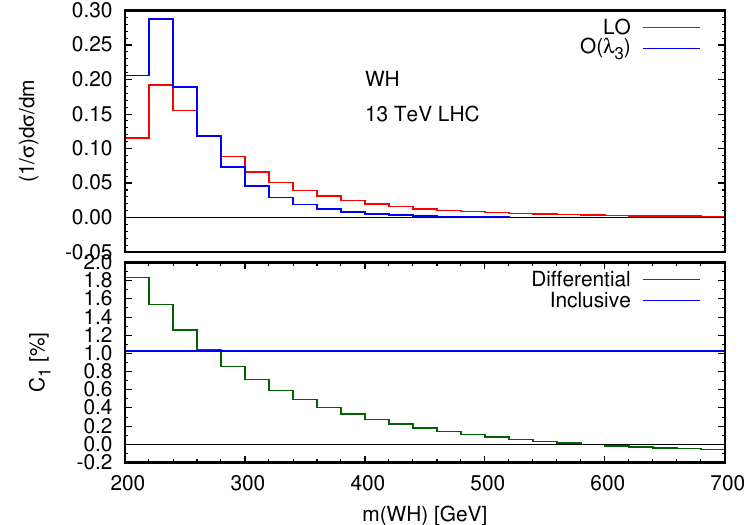}
\end{subfigure} 
\begin{subfigure}{0.49\linewidth}
  \includegraphics[width=7.8cm,clip]{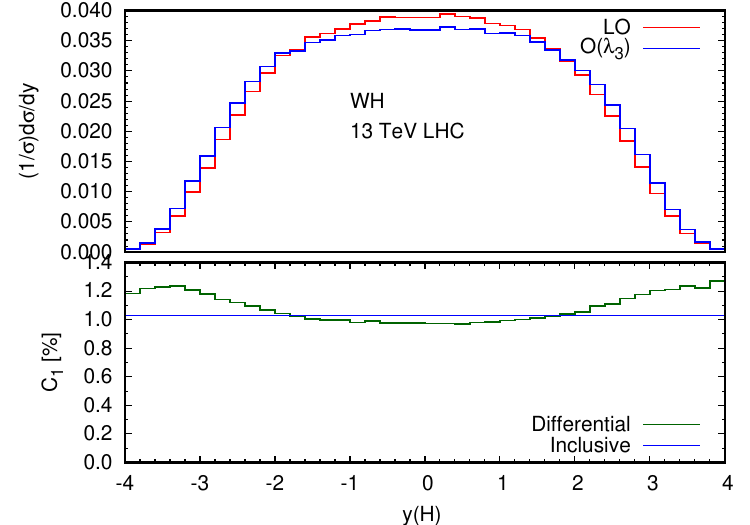}
\end{subfigure}  
\begin{subfigure}{0.49\linewidth}
  \includegraphics[width=7.8cm,clip]{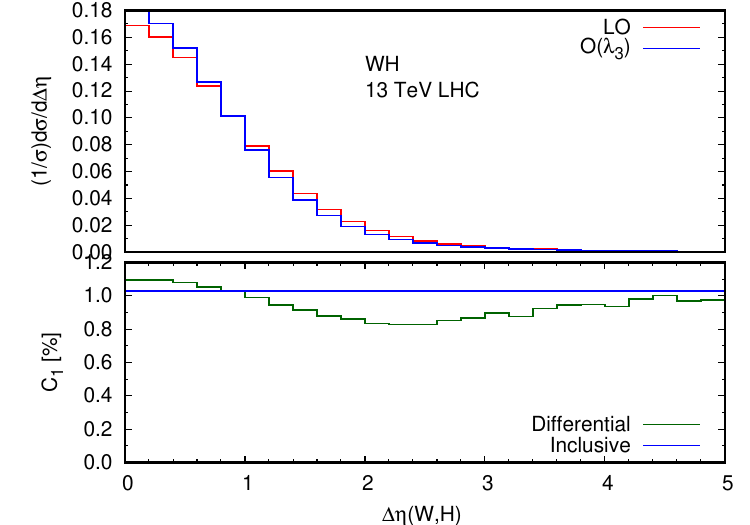}
\end{subfigure}  
\caption{ Effect of ${\mathcal O}(\lt)$ correction in $WH$ at 13 TeV LHC. Upper panel: 
         normalised distributions at LO (red) and at ${\mathcal O}(\lt)$ (blue). 
          Lower panel: $C_1$ at the differential (green) and inclusive (blue) level.} 
\label{fig:wh-lo-l3}       
\end{figure}
 In Figs.~\ref{fig:zh-lo-l3} and \ref{fig:wh-lo-l3} we show the differential $C_1$ for $ZH$ and $WH(W=W^+ , W^-)$, respectively. As discussed in refs~\cite{Degrassi:2016wml, Bizon:2016wgr} the main enhancements are present at threshold, where the interaction of the final-state vector and Higgs bosons via a Higgs propagator leads to a Sommerfeld enhancement due to the  non-relativistic regime. Indeed the shape of the ${\mathcal O}(\lt)$ corrections is quite different from the LO case for  $p_T(H)$ and $m(VH)$ distributions; the former are softer than the latter. For this reason, $C_1$ grows at threshold, where, however, the cross section is rather small. 
  In particular,  while $C_1$ in $ZH$ ($WH$)  is 1.19 (1.03)\% at the inclusive level, it grows up to,  {\it e.g.}, 
 2.3(1.8)\% for $m(ZH)$ at threshold, with the binning used in Figs. \ref{fig:zh-lo-l3}(\ref{fig:wh-lo-l3}). 
 Thus, in order to detect anomalous $\lt$ effects,
dedicated measurements close to threshold but with enough events, such as the region $p_T(H)<75$ GeV, would be desirable. 
 For $VH$ we also show $C_1$ for the rapidity $y(H)$ and the difference of the pseudo-rapidity of the $V$ and $H$ bosons
 $\Delta \eta(V,H)$. The latter is particularly interesting because $C_1$ is enhanced w.r.t. the inclusive case in the region corresponding to the largest cross section.
 
 We also looked at possible effects due to the $Z$ polarisation or, in other words, measurable via the angular distributions of $Z$ (and $H$) decay products. We did not see any enhancement or shape dependence for these distributions. Furthermore, one should bear in mind that also the loop-induced $gg\to HZ$ process gives a non-negligible amount of the NNLO cross section, order $\sim 1/6$ at 13 TeV. This process also has a dependence on $\lt$, but only at two-loop level, and should exhibit a shape dependence. However, this calculation is not technically feasible yet.
 
 \subsection {${\bf t{\bar t}H}$}
 \label{ttHsec}
 %
 %
 \begin{figure}[t]
\begin{subfigure}{0.5\linewidth}
  \includegraphics[width=7.8cm,clip]{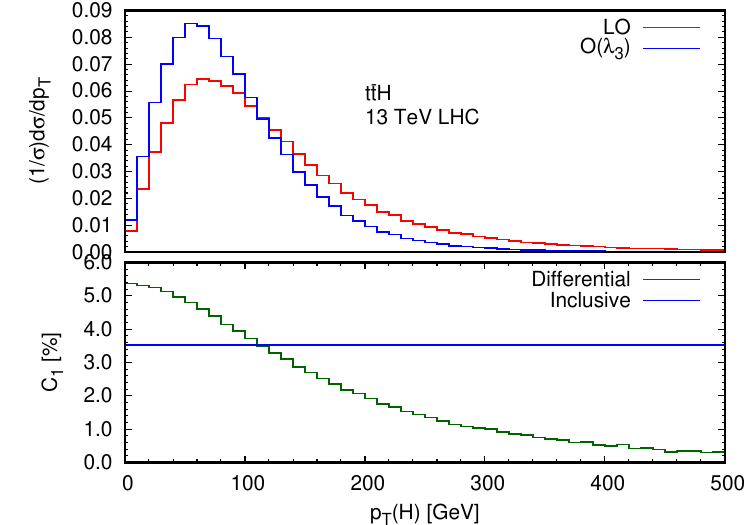}
\end{subfigure}
\begin{subfigure}{0.5\linewidth}
  \includegraphics[width=7.8cm,clip]{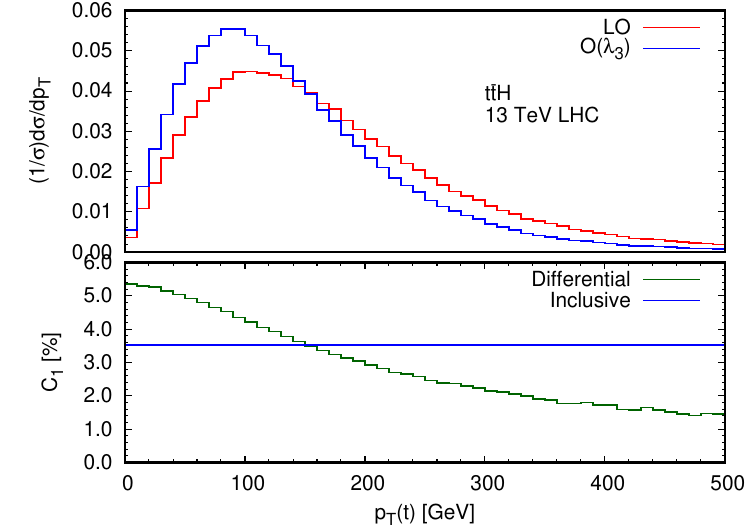}
\end{subfigure} 
\begin{subfigure}{0.5\linewidth}
  \includegraphics[width=7.8cm,clip]{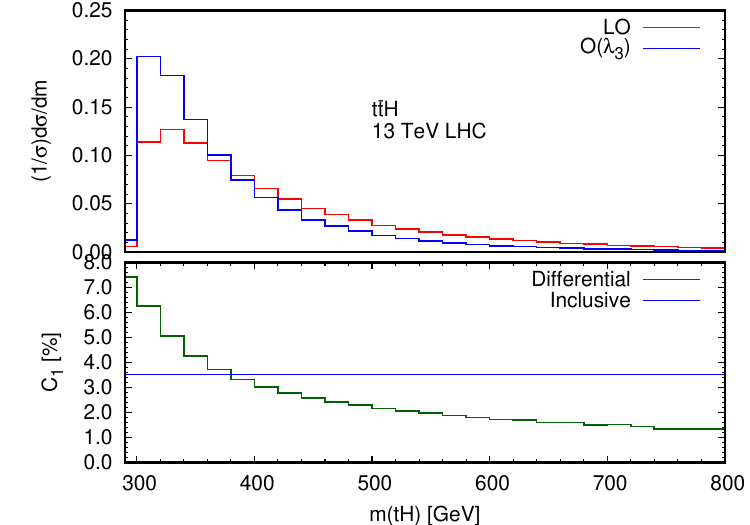}
\end{subfigure}
\begin{subfigure}{0.5\linewidth}
  \includegraphics[width=7.8cm,clip]{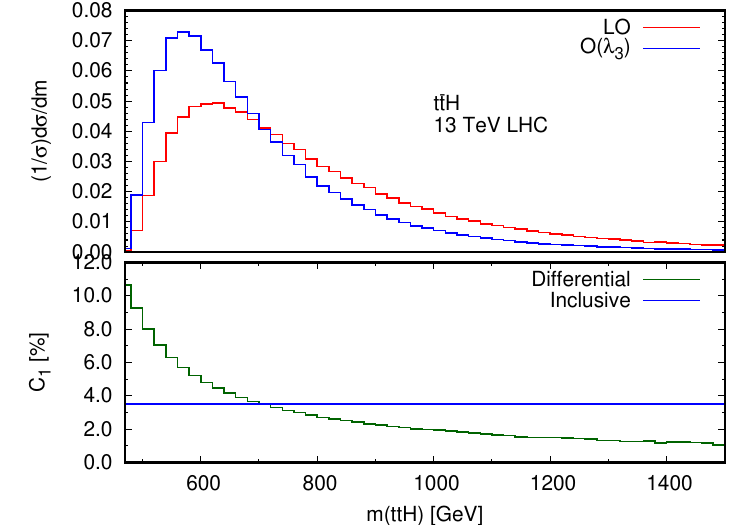}
\end{subfigure} 
\caption{ Effect of ${\mathcal O}(\lt)$ correction in $t \bar t H$ at 13 TeV LHC. Upper panel: 
         normalised distributions at LO (red) and at ${\mathcal O}(\lt)$ (blue). 
          Lower panel: $C_1$ at the differential (green) and inclusive (blue) level.} 
\label{fig:tth-lo-l3}       
\end{figure}
 Together with gluon-fusion production, the $t \bar t H$ channel plays a major role in providing information of the top-quark couplings to the Higgs. Its importance can be gauged by simply considering its weight in a global  $\kappa$-framework fit~\cite{Heinemeyer:2013tqa} or in the SMEFT framework~\cite{deFlorian:2016spz}. 
The same importance should be ascribed to this process also from the point of view of the sensitivity to $\lt$:  $C_1$ for $t \bar t H$ is the largest among all production channels and with the most significant kinematic dependence~\cite{Degrassi:2016wml}.  
In Fig.~\ref{fig:tth-lo-l3}, we show the most important kinematic distributions in 
 this channel. $C_1$ for total cross section is $3.52\%$ and 
 can increase up to  $\sim 5\%$ in $p_T$ distributions. Similarly, with the binning chosen in Fig.~\ref{fig:tth-lo-l3}, $C_1$ for the invariant mass distributions can be as large 
 as 10\% close to threshold, even though, once again, in the same region the cross section is suppressed
 by phase space. The origin of the large phase-space dependence of $C_1$ is again due to Sommerfeld enhancements in the threshold regions that are induced by interactions among the top (anti)quark and the Higgs boson. 
 \subsection{${\bf tHj}$} 
\label{sth}
 \begin{figure}[t]
\begin{subfigure}{0.5\linewidth}
  \includegraphics[width=8cm,clip]{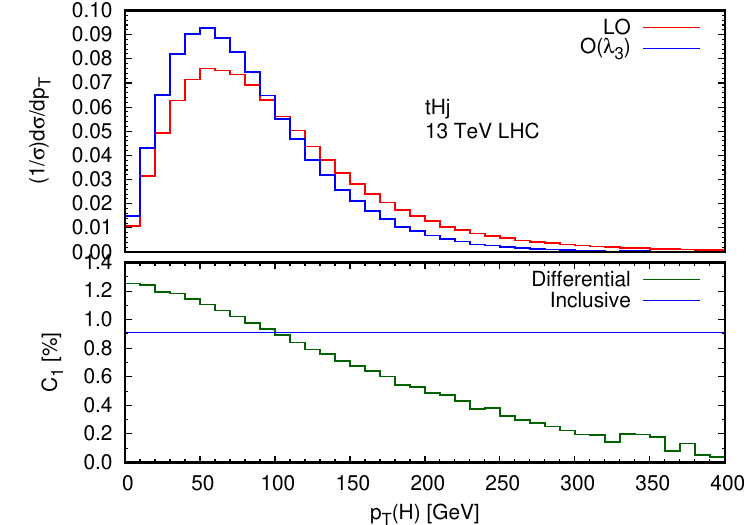}
\end{subfigure}
\begin{subfigure}{0.5\linewidth}
  \includegraphics[width=8cm,clip]{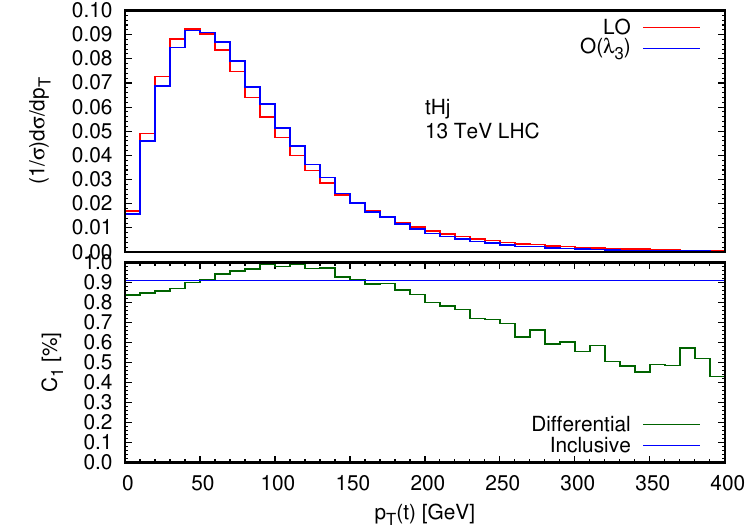}
\end{subfigure} 
\begin{subfigure}{0.5\linewidth}
  \includegraphics[width=8cm,clip]{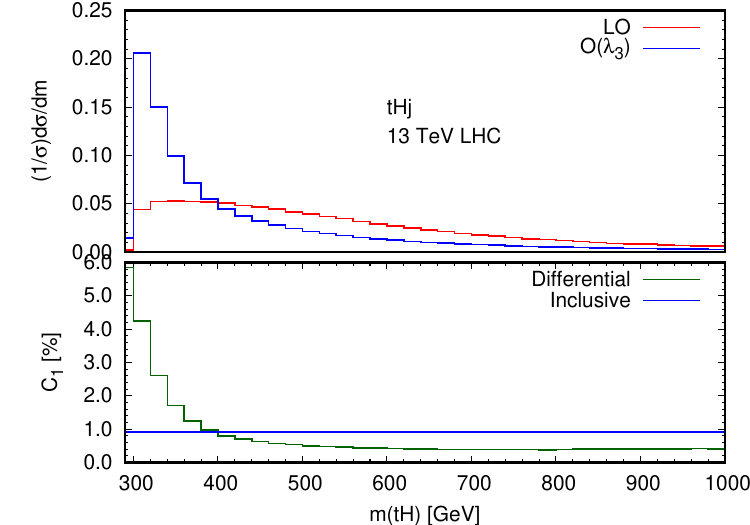}
\end{subfigure}
\begin{subfigure}{0.5\linewidth}
  \includegraphics[width=8cm,clip]{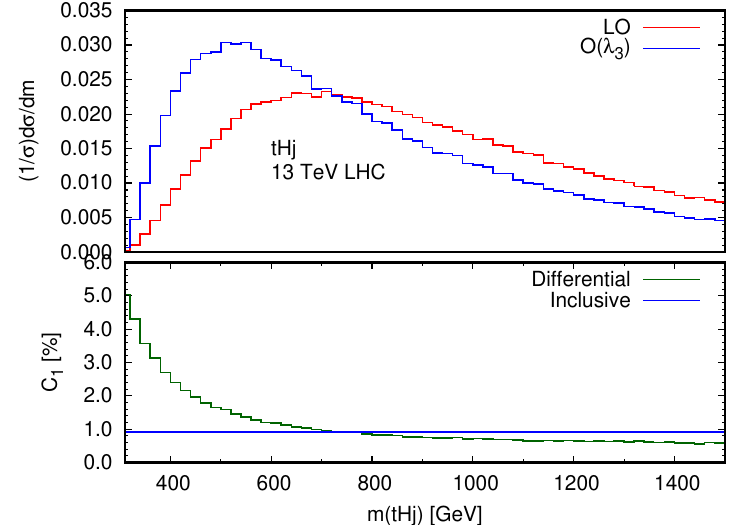}
\end{subfigure} 
\caption{ Effect of ${\mathcal O}(\lt)$ correction in $tHj$ at 13 TeV LHC. Upper panel: 
         normalised distributions at LO (red) and at ${\mathcal O}(\lt)$ (blue). 
          Lower panel: $C_1$ at the differential (green) and inclusive (blue) level.}
\label{fig:thj-lo-l3}       
\end{figure}
Although it is characterised by a rather small cross section at the LHC, single associate production of a Higgs with a single top is a particularly rich and interesting process, especially in searching for observables sensitive to relative phases among the Higgs couplings to fermions and bosons~\cite{Maltoni:2001hu,Biswas:2012bd,Farina:2012xp,Demartin:2015uha}. Naively, one would expect this process to have a sensitivity to the trilinear between that of VBF and $t \bar t H$; the $tHj$ process features a top quark in the final state as well as $W$ boson(s) in the propagators.  The contribution of one-loop diagrams featuring the Higgs self coupling to this process has not been considered in ref.~\cite{Degrassi:2016wml} for two main reasons. 
The first one was of phenomenological nature: in the SM this process   is barely observable at the Run II of the LHC. The second one is of technical nature, as the calculation needs a careful check of EW gauge invariance and UV finiteness, since a few subtleties, which are not present for the other processes discussed in this work, arise. We describe them in the following.

Similar to the case of the $H\rightarrow \gamma \gamma$ decay ~\cite{Gorbahn:2016uoy,Degrassi:2016wml}, Goldstone bosons appear in the Feynman diagrams contributing to the LO. Thus, $HGG$ as well $HHGG$ interactions are present in one-loop EW corrections. While the former is not modified by $(\Phi^\dag \Phi)^{n}$ effective operators, the latter is indeed modified ~\cite{Gorbahn:2016uoy,Degrassi:2016wml}. The calculation can be consistently performed in two different ways: either directly eliminating Goldstone bosons by employing the unitary gauge, as also done  for other quantities in refs.~\cite{Degrassi:2016wml, Degrassi:2017ucl}, or keeping track of $HHGG$ effects in the intermediate calculation steps, as we explain in the following and done in our calculation.

In a generic gauge, the on-shell renormalisation of the EW sector~\cite{Denner:1991kt} involves the counterterm for the Goldstone self-energy, which  depends on the Higgs tadpole counter term $\delta t$, which in turn depends on the trilinear coupling $\lt$. Therefore, if we only modify the value of $\lt$, the Goldstone self-energy counterterm receives a UV-divergent contribution proportional to ($\kappa_3-1$), which is not cancelled by any divergence from loop diagrams. Instead, if we consistently take into account the modification of the $HHGG$ vertex, loop diagrams featuring a seagull in the $G$ propagator are also present; they exactly cancel the UV-divergent  contribution proportional to $(\kappa_3-1)$ in the Goldstone self-energy counter term, leading to the same result one would obtain in the unitary gauge. Having understood this point, the calculation is straightforward and can be performed automatically in the Feynman gauge.

 In our results we include both $tHj$ and ${\bar t}Hj$ channels and we do not apply cuts on the jet, since the result is infrared finite. We find the $C_1$ for the total cross section is  about 0.91\%. In Fig.~\ref{fig:thj-lo-l3}, we show $C_1$ for kinematic distributions such as $p_T(H)$, $p_T(t)$, $m(tH)$ and $m(tHj)$. We note that unlike other variables  $p_T(t)$ does not decrease monotonically as we move from low to high $p_T$ values.  Near threshold $m(tH)$ displays a quite impressive difference in shape.
 \begin{figure}[t]
\begin{subfigure}{0.5\linewidth}
  \includegraphics[width=8cm,clip]{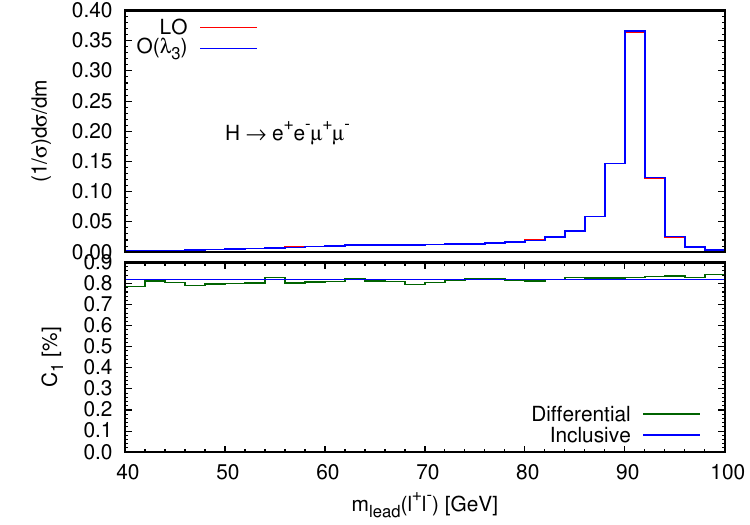}
\end{subfigure}
\begin{subfigure}{0.5\linewidth}
  \includegraphics[width=8cm,clip]{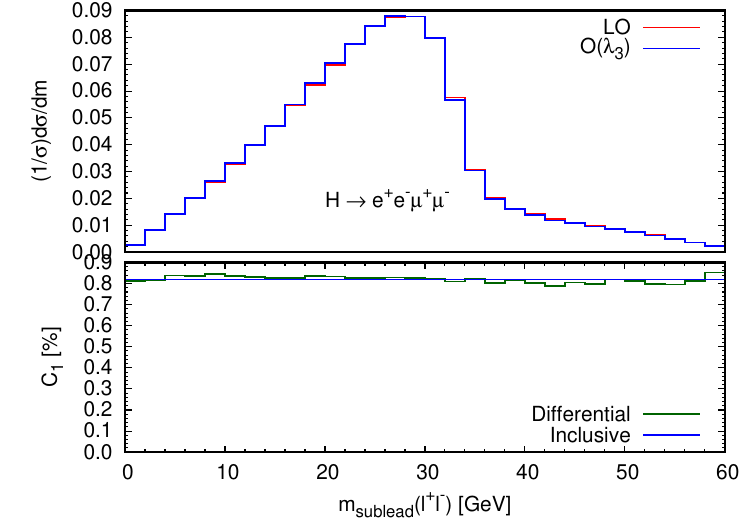}
\end{subfigure} 
\caption{ Leading (left) and subleading (right) OSSF lepton pair invariant mass distributions in 
$H \to e^+e^-\mu^+\mu^-$. 
Upper panel: normalised LO (red) and ${\mathcal O}(\lt)$ (blue) distributions. Lower panel: $C_1$ for differential (green) 
and total decay width (blue).}   
\label{fig:h4l}       
\end{figure}
\subsection {${\bf H \to 4\ell}$}
The Higgs decay into four fermions is the only Higgs decay channel with 
nontrivial final state kinematics. Moreover, it is the only one where {\it a priori} also $C_1$ can have a shape dependence. Indeed, all the other decays correspond to a $1\to 2$ process, and since the $H$ boson is a scalar, there is not a preferred direction in its reference frame. In the previous study \cite{Degrassi:2016wml} the $C_1$ for $H \to ZZ^*$ decay 
was calculated to be 0.83$\%$. Although, the full off-shell configuration was taken into account, possible angles between the decay products were not analysed. Using the form factor code mentioned above we calculate $C_1$ for $H \to e^+e^-\mu^+\mu^-$ channel.  
We analysed $C_1$ for many observables involving the four leptons, but we found that it has in general almost no kinematic dependence. 
As an example, in Fig.~\ref{fig:h4l}, we display $C_1$ for leading and sub-leading lepton pair invariant 
masses. 
Since the Higgs boson interactions with the final-state fermions are negligible, this result can be extended to all the other decays into four leptons and in general into four fermions.
\clearpage

 \section{Anomalous trilinear effects and the NLO electroweak corrections}\label{sec:nlo-ew}
 %
\begin{figure}[t]
\begin{subfigure}{0.5\linewidth}
  \includegraphics[width=8cm,clip]{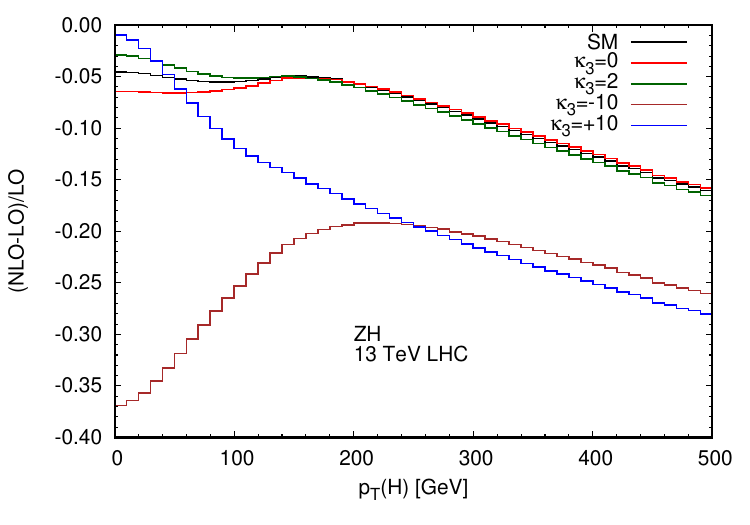}\caption{}\label{fig:zh-nlo-sm-pt}
\end{subfigure} 
\begin{subfigure}{0.5\linewidth}
  \includegraphics[width=8cm,clip]{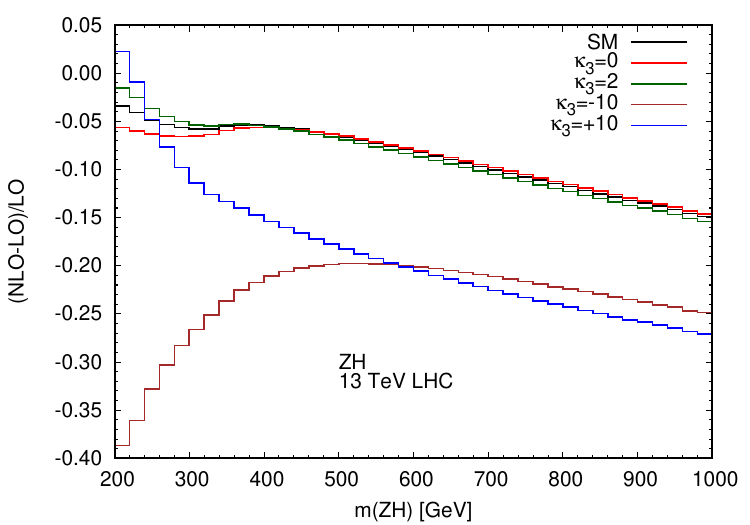}\caption{}\label{fig:zh-nlo-sm-mzh}
\end{subfigure}  
\begin{subfigure}{0.5\linewidth}
  \includegraphics[width=8cm,clip]{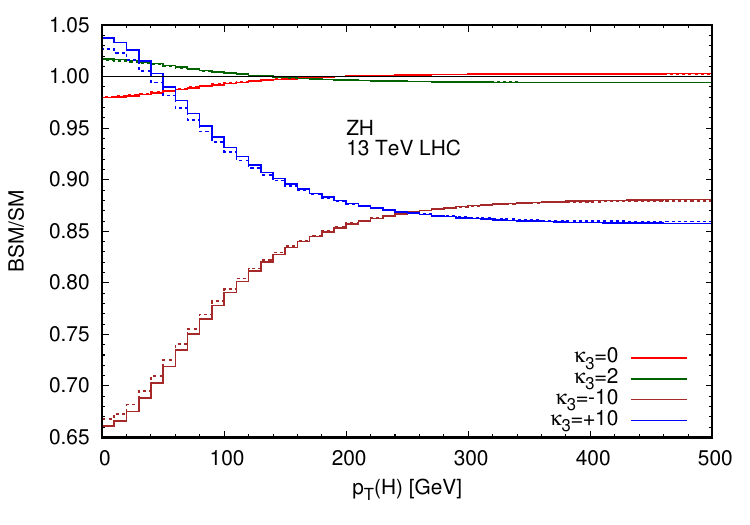}\caption{}
\end{subfigure}  
\begin{subfigure}{0.5\linewidth}
  \includegraphics[width=8cm,clip]{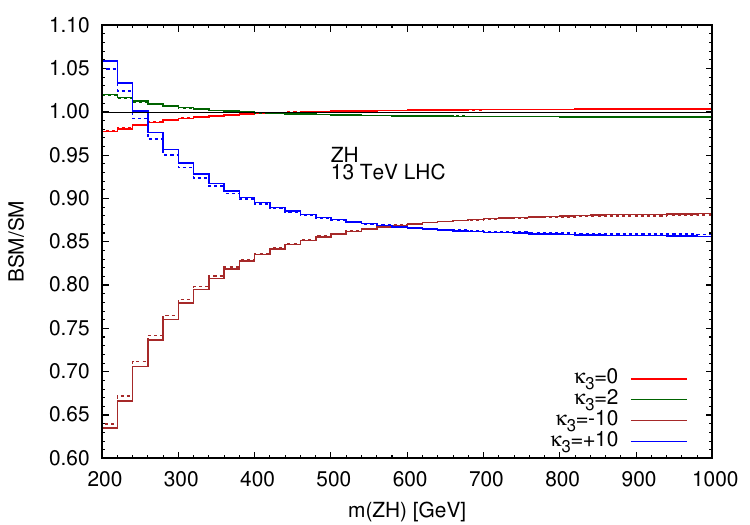}\caption{}
\end{subfigure}
\caption{ $p_T(H)$ (left) and $m(ZH)$ (right) distributions for $ZH$. Upper plots: $(\sigma_{\rm NLO}^{\rm BSM}-\sigma_{\rm LO})/\sigma_{\rm LO}$ ratio for different values of $\kt$. Lower plots: comparison of BSM/SM ratio including or not NLO EW corrections for different values of $\kt$.}
\label{fig:zh-nlo}       
\end{figure}
%
 %
\begin{figure}[t]
\begin{subfigure}{0.5\linewidth}
  \includegraphics[width=8cm,clip]{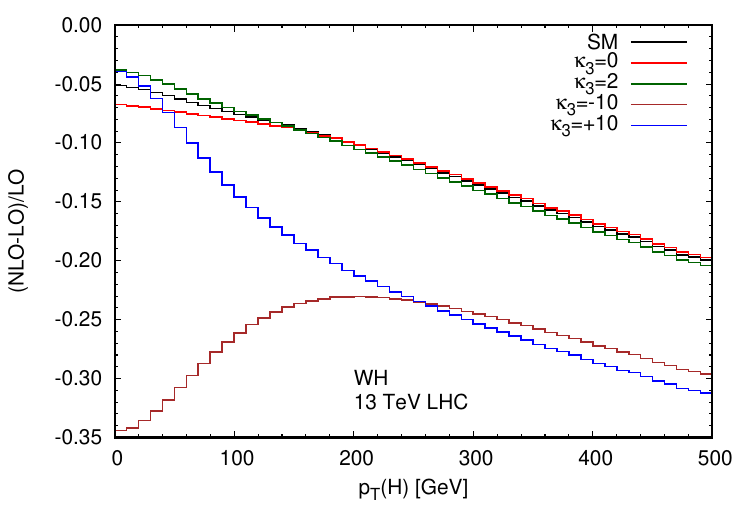}\caption{}
\end{subfigure} 
\begin{subfigure}{0.5\linewidth}
  \includegraphics[width=8cm,clip]{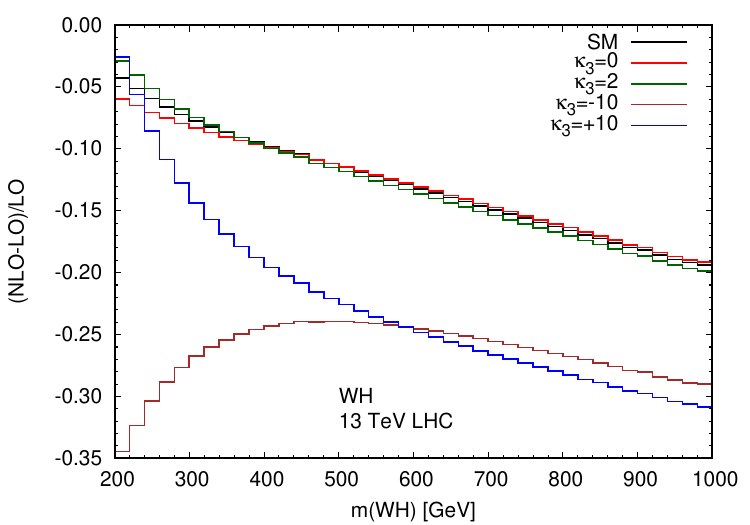}\caption{}
\end{subfigure} 
\begin{subfigure}{0.5\linewidth}
  \includegraphics[width=8cm,clip]{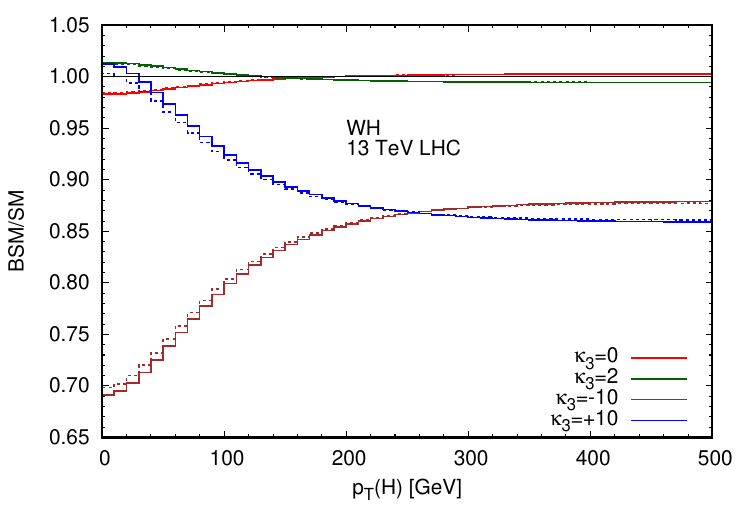}\caption{}
\end{subfigure}  
\begin{subfigure}{0.5\linewidth}
  \includegraphics[width=8cm,clip]{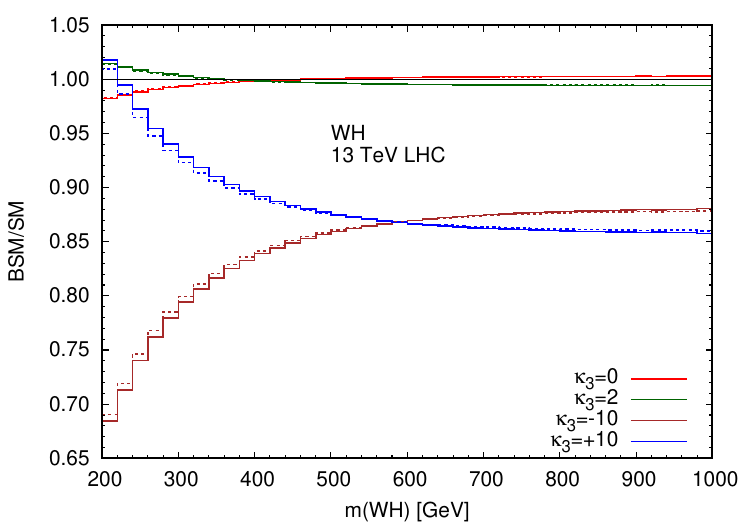}\caption{}
\end{subfigure}
\caption{ $p_T(H)$ (left) and $m(WH)$ (right) distributions for $WH$. Upper plots: $(\sigma_{\rm NLO}^{\rm BSM}-\sigma_{\rm LO})/\sigma_{\rm LO}$ ratio for different values of $\kt$. Lower plots: comparison of BSM/SM ratio including or not NLO EW corrections for different values of $\kt$.}
\label{fig:wh-nlo}       
\end{figure}
%
%
\begin{figure}[t]
\begin{subfigure}{0.5\linewidth}
  \includegraphics[width=8cm,clip]{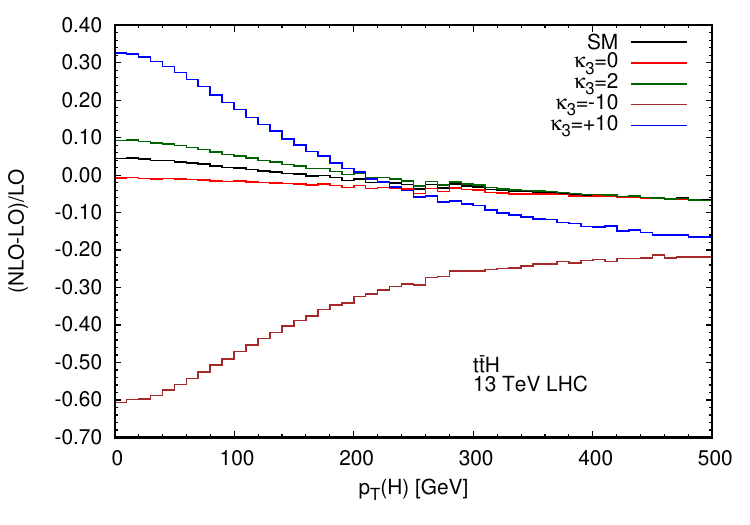}\caption{}
\end{subfigure} 
\begin{subfigure}{0.5\linewidth}
  \includegraphics[width=8cm,clip]{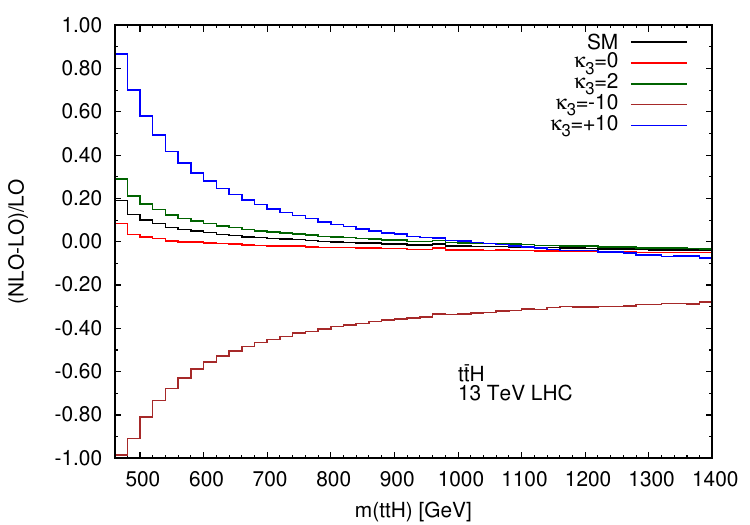}\caption{}
\end{subfigure}  
\begin{subfigure}{0.5\linewidth}
  \includegraphics[width=8cm,clip]{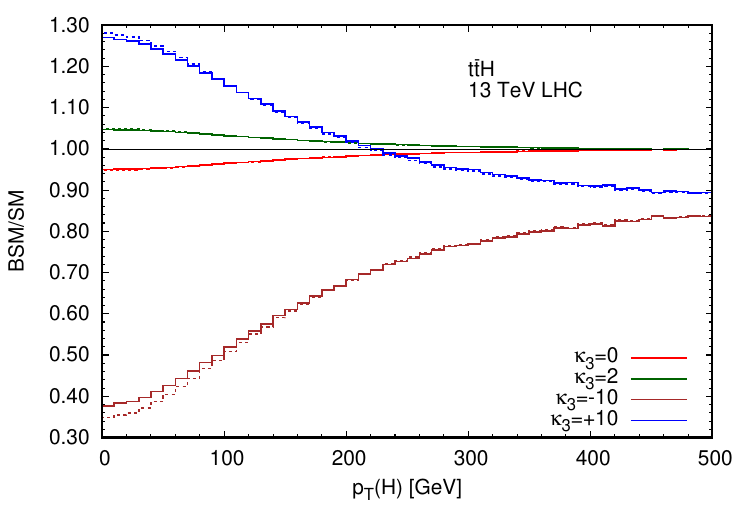}\caption{}
\end{subfigure}  
\begin{subfigure}{0.5\linewidth}
  \includegraphics[width=8cm,clip]{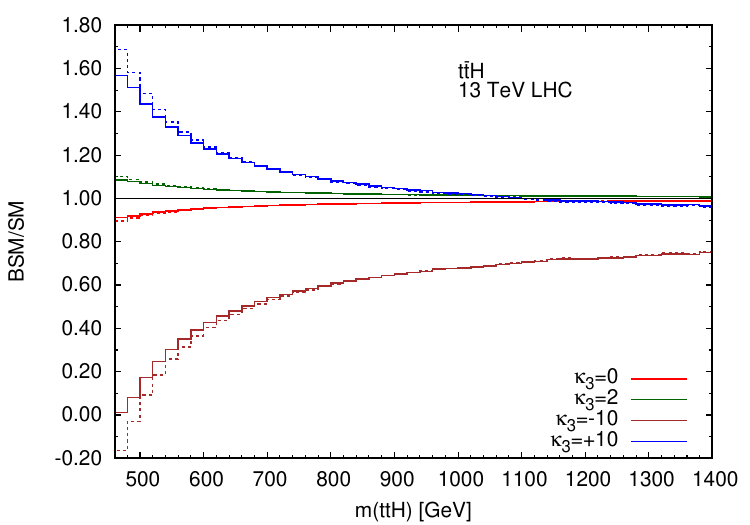}\caption{}
\end{subfigure}
\caption{ $p_T(H)$ (left) and $m(t \bar t H)$ (right) distributions for $t \bar t H$. Upper plots: $(\sigma_{\rm NLO}^{\rm BSM}-\sigma_{\rm LO})/\sigma_{\rm LO}$ ratio for different values of $\kt$. Lower plots: comparison of BSM/SM ratio including or not NLO EW corrections for different values of $\kt$.}
\label{fig:tth-nlo}       
\end{figure}
\begin{figure}[t]
   \centering
 \includegraphics{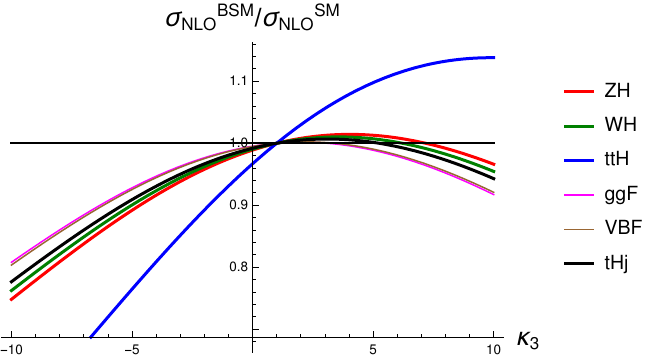}
   \caption{ $\sigma_{\rm NLO}^{\rm BSM}/\sigma_{\rm NLO}^{\rm SM}$ as a function 
   of $\kt$ for $gg$F (purple), VBF (brown), $WH$ (green), $ZH$ (red), $t \bar t H$ (blue),  and $tHj$ (black) 
   at 13 TeV LHC. }
   \label{fig:nlo-ratio}
 \end{figure}
The set of one-loop corrections to single Higgs production and decays involving the trilinear 
 Higgs self coupling  is gauge invariant and finite. However, in the SM, performing a perturbative expansion in power of $\alpha_s$ and $\alpha$, other contributions are present at the same order of accuracy. In other words, $\lt$-induced effects at one-loop level should be considered as a gauge-invariant and finite subset of the complete NLO EW corrections, which also include  effects form virtual $W,Z$ and photons as well as real emissions contributions.~\footnote{In the EW sector of the SM all the interactions are determined by the mass of the fermions, $m_H$ and three additional parameters, which are typically $m_W$, $m_Z$ and $\alpha$ or $G_\mu$. In general, it is not possible to alter at NLO EW accuracy a derived quantity, such as $\lambda=m_H^2/(2 v^2)$, without spoiling the renormalisability of the theory. The special case of $\lt$ in single Higgs production at one loop has been discussed in detail in refs.~\cite{Degrassi:2016wml} and \cite{Degrassi:2017ucl}.   }  
 
 As shown in the previous section, the possibility of measuring anomalous $\lt$ effects via precise predictions in single Higgs production relies both on the precision of experimental measurements and SM theory predictions. In particular, regarding the theory accuracy, while it is reasonable to assume that QCD corrections  in general factorise $\lt$ effects, as explained in sec.~\ref{sec:c1-calc},  this is in general not true for NLO EW corrections. 
 The purpose of this section is to provide a consistent extension of the master formula in eq.~\eqref{master} that includes also NLO EW corrections and to investigate  their  impact in the determination of  anomalous $\lt$ effects. All the calculations of the NLO EW corrections in the SM, with the exception of $gg$F taken from \cite{deFlorian:2016spz}, are performed in a completely automated approach via an extension of the {\sc MadGraph5\_aMC@NLO} framework that has already been used and validated in refs.~\cite{Frixione:2014qaa,Frixione:2015zaa, Badger:2016bpw, Pagani:2016caq, Frederix:2016ost, Czakon:2017wor}. Concerning the renormalisation,  we use the $G_\mu$-scheme, consistently with the input parameters listed in sec.~\ref{sec:c1-dist}.
 
  At the differential level we limit ourselves to the study of the $VH$ and $t \bar t H$ processes, where the $C_1$ dependence on the kinematics is large. We have also computed the differential EW corrections to the $tHj$ production channel, but we do not report plots here, as its phenomenological relevance  will be marginal at 13 TeV LHC for our purposes. The differential case is of particular interest since EW corrections in the SM, due to the Sudakov logarithms,  are  large in the boosted regime, {\it i.e.}, exactly in the opposite phase-space region where $\lt$-induced effects are sizeable, the production threshold, as already discussed before.
  
The master formula in eq.~\ref{master} can be improved including NLO EW correction in the following way 
 \begin{equation}
  \Sigma_{\rm NLO}^{\rm BSM} = Z_H^{\rm BSM}\Big[ \Sigma_{\rm LO}\left(1+\kt C_1+\delta Z_H+\delta_{\rm EW}\big|_{{\lambda}_3=0}\right)  \Big]\,, \label{predNLOBSM}
 \end{equation}
where $\delta_{\rm EW}\big|_{{\lambda}_3=0}$ represents the part of the NLO EW $K$-factor in the SM
 \begin{equation}
  K_{\rm EW} \equiv \frac{ \Sigma_{\rm NLO}^{\rm SM} }{ \Sigma_{\rm LO} }\,, \label{KEW}
 \end{equation} 
   that does not depend on $\lt$, namely~\footnote{Here, in order to keep the notation simple, with the symbol $\delta Z_H$ we still refer to only the $\lt$ contributions to the Higgs wave-function counterterm. Thus, $\delta_{\rm EW}\big|_{{\lambda}_3=0}$ contains further contributions  to the Higgs wave-function counterterm that do not depend on $\lt$.},
 \begin{equation}
  \delta_{\rm EW}\big|_{{\lambda}_3=0} \equiv K_{\rm EW} - 1- C_1- \delta Z_H\,.
 \end{equation} 
 In eq.~\eqref{KEW}, $ \Sigma_{\rm NLO}^{\rm SM}$ stands for the observable $ \Sigma$ at LO + NLO EW accuracy. Thus, in the limit $\lt \rightarrow 1$, $\Sigma_{\rm NLO}^{\rm BSM} \rightarrow \Sigma_{\rm NLO}^{\rm SM}$. As can be noted, the $Z_H^{\rm BSM}$ term factorises the NLO EW contributions in the SM, while $C_1$ does not. Indeed, in general, EW loop corrections on top of $\lt$-induced effects need a dedicated two-loop calculation and a full-fledged EFT approach in order to obtain UV-finite results; only the $Z_H^{\rm BSM}$ contribution is completely model-independent and factorises the NLO EW corrections in the SM. However, it is worth to note that, assuming factorisation also for $C_1$ contributions, terms of the order $\kappa_3 C_1 \times \delta_{\rm EW}\big|_{{\lambda}_3=0} $ would be anyway negligible, since either $ \delta_{\rm EW}\big|_{{\lambda}_3=0} $ (Sudakov logarithms in the boosted regime) or   
$C_1$ (Sommerfeld enhancement in the threshold region) is sizeable, but never both of them at the same time. This will be clear in the differential plots we display in the following. 
 \begin{table}[b]
\begin{center}
\begin{tabular}{|c |c| c| c| c| c| c|}
\hline
 Channels &$gg$F &VBF &$ZH$ & $WH$  &$ttH$ &$tHj$   \\ 
\hline
 $K_{\rm EW}$ &1.049 &0.932 &0.947 &0.93  &1.014 &0.95   \\
\hline
\end{tabular}
\caption{NLO EW K-factors for different production channels.}
\label{tab:kfactor-xs}
\end{center}
\end{table} 
  
The EW $K$-factor at the inclusive level can be found for all processes in Tab.~\ref{tab:kfactor-xs}, while relevant differential results for $ZH$, $WH$ and $t \bar t H$ are displayed in Figs.~\ref{fig:zh-nlo}, \ref{fig:wh-nlo} and \ref{fig:tth-nlo}, respectively. In each figure, plots on the left show the $p_T(H)$ distributions, while plots on the right those for the invariant mass of the final state. In the upper plots we display the ratio $(\sigma_{\rm NLO}^{\rm BSM}-\sigma_{\rm LO})/\sigma_{\rm LO}$ for different values of $\lt$, (-10,0,1,2,10). In practice, the case $\lt=1$, directly denoted as SM in the plots, corresponds to the differential $(K_{\rm EW}-1)$ in the SM. The lower plots display the ratio $\sigma_{\rm NLO}^{\rm BSM}/\sigma_{\rm NLO}$ (solid lines) and the term $1 +\delta\Sigma_{\kt}$ (dashed lines) for different values of $\lt$, (-10,0,1,2,10). In practice, the former is our prediction at NLO EW accuracy for the signal strengths $\mu_{i}$  that will enter in the fits of the next section\footnote{The signal strengths $\mu_{i}$  is better defined afterwards  in eq.~\eqref{mui}. In this section $\kappa_i=1$.}, the latter is the definition at LO used in the previous works.

First, we comment on the shape of the NLO EW corrections in the upper plots. The general trend in the SM is characterised by large negative Sudakov logarithms in the tails, especially for  $p_T(H)$ in $VH$ production, and positive corrections in the threshold region, especially for the invariant mass distributions and in general $t \bar t H$. The latter are precisely the effects due to $C_1$. Thus, changing the value of $\lt$, the shape of the $\sigma_{\rm NLO}^{\rm BSM}/\sigma_{\rm LO}$ ratio is highly affected in the threshold region, while it is not deformed in the tail. On the other hand, the change induced by $\lt$ on $Z_H^{\rm BSM}$ results in a constant shift in the tail of the distributions. The small bump  around $p_T(H)\sim m_t$ and $m(ZH)\sim 2m_t$  in $ZH$ production is simply due to the $t{\bar t}$ threshold in diagrams involving a top-quark loop.

By looking at the upper plots in Figs.~\ref{fig:zh-nlo}, \ref{fig:wh-nlo} and \ref{fig:tth-nlo} is evident that EW corrections have to be included in order to correctly identify anomalous $\lt$ effects. On the other hand, NLO EW corrections do not largely affect the value of the signal strengths, {\it i.e.} the ratio of the BSM and SM prediction. This fact can be seen in the lower plots, where we display $\sigma_{\rm NLO}^{\rm BSM}/\sigma_{\rm NLO}$ (solid lines) and $1 +\delta\Sigma_{\kt}$ (dashed lines), which indeed corresponds to the aforementioned ratio with or without NLO EW corrections both in the numerator and denominator.
\footnote{Actually, the ratio without NLO EW corrections should be $\sigma_{\rm \lt}^{\rm BSM}/\sigma_{\rm \lt}^{\rm SM}$, but its difference with $1 +\delta\Sigma_{\kt}$, which is used in previous works and more useful for a direct comparison, is negligible.} 
Solid and dashed lines are in general very close, especially for small values of $\kt$.  It is interesting to note that for very small values of $m(t \bar t H)$ a value $\kl=-10$ is leading to corrections that are negative and larger in absolute value than the LO. This is due to the very large $C_1$ (see Fig.~\ref{fig:tth-lo-l3}) and points to the necessity of including higher order $\lt$-induced effects for large values of $\kt$ and for this specific phase-space region.

For the case of total cross sections, we plot the ratio $\sigma_{\rm NLO}^{\rm BSM}/\sigma_{\rm NLO}^{\rm SM}$ 
as a function of $\kt$ in the range [-10, 10] for all the single Higgs production processes, 
including also $tHj$. Differences with the corresponding $1 +\delta\Sigma_{\kt}$ ratios, which have been presented in ref.~\cite{Degrassi:2016wml}, are hardly visible and thus we do not show them here.

In conclusion, constraints on $\lt$ from a global fit based on the value of the signal strengths $\mu_i$ at inclusive \cite{Degrassi:2016wml} or differential level \cite{Bizon:2016wgr} will not be affected by NLO EW corrections. On the other hand, in the experimental analyses EW corrections have to be taken into account, especially at the differential level, for  the determination of the value of the signal strengths, which is in general important for any BSM study and not peculiar for our case. 
 \section{Constraining $\kappa_3$ through a global fit}\label{sec:global-fit}
\label{sec:global-fit}
In this section we discuss the role of differential distributions in the determination of $\lt$ via  single-Higgs production and decays measurements.
The aim of this section is threefold. First, we show that the interplay of theory and experimental uncertainties have a large impact in the determination of the constraints on $\lt$,  especially when differential information is exploited. Second, we discuss how bounds on $\lt$ are affected by the presence of additional anomalous couplings in the fit. In particular, we progressively lift the assumption that the Higgs couplings to the top quark  and to vector bosons are SM-like. Third, we include the EW effects discussed in the previous section in the fit analysis, providing consistent formulas for repeating the fit in conjunction with additional Higgs anomalous couplings. The reader that is only interested in the results can go directly to sec.~\ref{secresults} and skip sec.~\ref{secformulas}, where  formulas are given and a few technical details are discussed.

\subsection{Combined parameterisation of $\kappa_3$, $\kappa_t$ and $\kappa_V$ effects  }
\label{secformulas}
In order to parametrise the Higgs anomalous couplings to the top quark  and to the vector bosons we use the coupling modifiers $\kappa_t$ and $\kappa_V$, respectively (see ref.~\cite{Khachatryan:2016vau} for definitions). We are interested in how additional BSM effects entering at LO may alter the determination of $\kt$ and the relevance of differential measurements. The choice of the ($\kt, \kappa_t, \kappa_V$)  kappa framework is driven by simplicity; our main purpose is adding new degrees of freedom in the fit and identify in which configurations the differential information may be particularly relevant. On the other hand, while the cases of new physics entering only via $\kt$ and a very general EFT parametrisation (10 independent parameters) have already been explored~\cite{DiVita:2017eyz}, simplified intermediate parameterisations have not been considered yet. These parameterisations, such as the one used here,  may be useful to identify relevant scenarios for which the determination of $\kt$ is feasible.~\footnote{ In fact, the analysis carried here is a particular choice of  two (linear combinations) of the 10 Wilson coefficients identified in ref.~\cite{DiVita:2017eyz}: $\kappa_t$ is related to $\delta y_t$ and $\kappa_V$ to $c_W=c_Z$.}

First of all we extended the framework and the notation introduced in ref.~\cite{Degrassi:2016wml} in order to take into account in the fit differential information, EW corrections in the production, and $\kappa_t$ and $\kappa_V$ dependence. The experimental inputs entering the fit are the signal strengths, which
are defined for any particular combination $i \to H \to f$ of production and decay
channel  as 
\begin{equation}
\mu_i^f\equiv \mu_i \times \mu^f =
\frac{\sigma(i)}{\sigma(i)^{\rm SM}} \times \frac{{\rm BR}(f)}{{\rm BR}^{\rm
    SM}(f)}~.
\label{signalstre}
\end{equation}
In eq.~\eqref{signalstre}, the quantities $\mu_i$ and $\mu^f$ are  the 
production cross sections $\sigma(i)$ 
($i=$ $gg$F, VBF, $WH$, $ZH$, $t \bar tH$) and the ${\rm BR}(f)$ $(f= \gamma\gamma, VV^*,ff)$ divided by their SM values, respectively. Assuming on-shell
production, the product $\mu_i \times \mu^f$ is the measured rate for the $i \to H \to f$ process 
divided by the corresponding SM prediction. This is valid also for differential distributions involving the reconstructed momentum of the Higgs boson. For simplicity, in the following we will refer with the symbol $\sigma$ to both total cross sections or (bins in) differential distributions.

The signal strength productions $ \mu_i$ are given by
  \begin{equation}
    \mu_i= \frac{\sigma_i^{\rm BSM}}{\sigma_i^{\rm SM}} = 1+  \delta \mu_i(\kappa_3) + Z_H^{\rm BSM} (\kappa_i^2-1)\,, \label{mui}\\
 \end{equation}
 where 
 $\kappa_{gg \rm F} = \kappa_{t \bar t H} =\kappa_t$ and 
 $\kappa_{VH} = \kappa_{\rm VBF} =\kappa_V$
 and the effect of $\kt$ is parameterised via the quantity $\delta\mu_i(\kappa_3)$.  In presence of NLO EW corrections $\delta\mu_i(\kt) $ is given by
 \begin{equation}
     \delta\mu_i(\kt) =  \frac{\sigma_{\rm NLO}^{\rm BSM}(i)}{\sigma_{\rm NLO}^{\rm SM}(i)}	-1  
        =  Z_H^{\rm BSM} \left[ 1 + \frac{ (\kappa_3-1)C_1^i }{ K_{\rm EW}(i)} \right]-1\, ,
\label{eq:mu-production}
 \end{equation}
where we have explicitly shown which quantities depend on the specific production process $i$. 
For differential distributions, differential  $K_{\rm EW}$ have to be used. 
As can be noted, we did not include $\kappa_t$ and $\kappa_V$ effects entering at 
one loop. As we will see in the results of the fit, we are going to probe deviations at the  
percent level in $\kappa_t$ and $\kappa_V$. Thus, $\kappa_t^2 \kt$ and $\kappa_V^2 \kt$ effects are negligible for our purposes.  On the other hand, terms of order $\kappa_t^2 \kt^2$ and $\kappa_V^2 \kt^2$ may be more important and in fact those from the Higgs-wave-function can be consistently resummed; they are included in eq.~\eqref{mui} via the $Z_H^{\rm BSM} \kappa_i^2$ term.~\footnote{Eq.~\eqref{eq:mu-production} can be in principle generalised to an EFT framework. In that case, EW corrections can be performed also on top of new physics effects entering at the tree level as well as $\kt$-induced correction. However, the latter involves non-trivial higher-dimensional corrections.} However, unless differently specified, we verified that also the inclusion of the $\kappa_t^2 \kt^2$ and $\kappa_V^2 \kt^2$ contributions has a negligible impact in the results presented in the following. It is also important to note that any further $\kappa_t $ or $\kappa_V$ dependence that may be introduced by NLO EW corrections, on top of those already present at LO, is negligible. Indeed, NLO EW corrections are {\it per se} at the percent level and their anomalous $\kappa_t$ component would be of the order of few percents of the corrections themselves. Thus, these kinds of effects, which similarly to those of order $\kappa_t^2 \kt^2$ and $\kappa_V^2 \kt^2$ can actually be calculated only in an EFT framework, are  expected to be of the order $\sim\alpha (\kappa_i -1)$ and therefore at the permille level or even smaller in our analysis. For this reason, we can safely ignore them.

Similarly, the signal strength $\mu_f$ for the Higgs decays $H \to f$ is given by 
 \begin{equation}
\mu_f = \frac{{\rm BR}^{\rm BSM}(f)}{{\rm BR}^{\rm SM}(f)} 
      =  \frac{\Gamma^{\rm BSM}(f)}{\Gamma^{\rm SM}(f)} \frac{\Gamma_H^{\rm SM}}{\Gamma_H^{\rm BSM}}\, . \label{muf}
 \end{equation}
 NLO EW corrections in Higgs decays are small at inclusive level, therefore we can safely ignore them. The partial 
 decay width in a given channel is given by 
 \begin{equation}
  \Gamma^{\rm BSM}(f) = Z_H ( \kappa_f^2 + \kt C_1^f )\Gamma_{ \rm LO}^{\rm SM}(f)\, ,  \label{gammamod}
 \end{equation}
 where $\Gamma_{ \rm LO}^{\rm SM}$ is the total width at LO in the SM.  The SM widths $\Gamma^{\rm SM}(f)$ in eq.~\eqref{muf}
 can be obtained by setting $\kt=\kappa_f=1$ in $\Gamma^{\rm BSM}(f)$.  In order to ensure that the contribution of the Higgs wave-function renormalisation does not affect the branching ratios, in this case we resummed also the SM part ($ Z_H = 1/(1-\kt^2 \delta Z_H)$) and factorise it to the $\kappa_f^2$ dependence, as done in ref.~\cite{Degrassi:2016wml} for the LO analysis.  
 For the $\gamma\gamma$ decay channel $\kappa_{\gamma\gamma}$ depends on $\kappa_t$ and 
 $\kappa_V$%
 ~\footnote{The relevant expression is \begin{equation}
                      \kappa_{\gamma\gamma} = 
                      \frac{|\kappa_V A_1(\tau_W)+ \kappa_t \frac{4}{3}A_{1/2}(\tau_t)|}{|A_1(\tau_W)+ \frac{4}{3}A_{1/2}(\tau_t) |}\,, \nonumber
                     \end{equation}
                     where the functions $A_1(\tau_W)$   and  $A_{1/2}(\tau_t)$ are defined in Ref.~\cite{Djouadi:2005gi}.}
, $\kappa_{VV^*}= \kappa_V$ and $\kappa_{ff} = 1$. Using 
 eq.~\ref{gammamod} the signal strength for the decay becomes 
  \begin{eqnarray}
\mu_f =&& 
\frac{\kappa_f^2  +\kt C_1^f }{\sum_j {\rm BR}^{\rm SM}_{\rm LO}(j) [\kappa_j^2  +\kt C_1^j]}
\frac{1 + \sum_j {\rm BR}^{\rm SM}_{\rm LO}(j)  C_1^j}{1+C_1^f} \\
      \simeq && \frac{\kappa_f^2  +(\kt-1) C_1^f }{\sum_j {\rm BR}^{\rm SM}(j) [\kappa_j^2  +(\kt-1) C_1^j]  } \, ,\label{mufformula}
 \end{eqnarray}
where in the last step we have assumed  that $C_1$ is small, which is indeed true for the decay channels.

\subsection{Results: comparison of differential and inclusive information in different scenarios}
\label{secresults}
   \begin{figure}[t]
  \begin{subfigure}{0.5\linewidth}
  \includegraphics[width=8cm,clip]{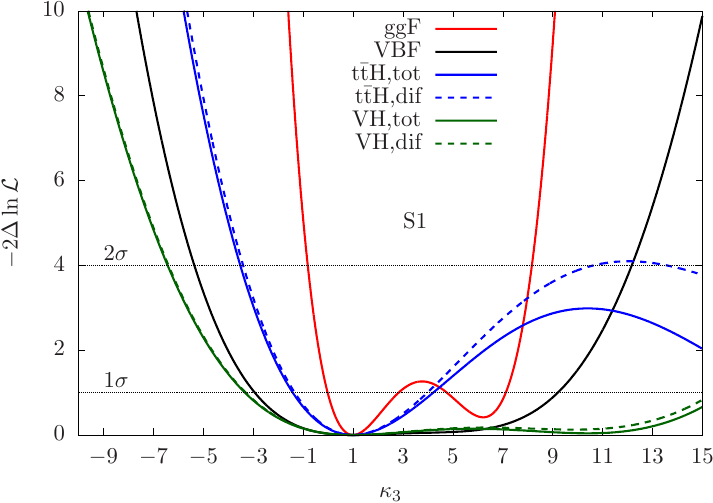}\caption{}\label{fig:fit-k3-each}
  \end{subfigure}
    \begin{subfigure}{0.5\linewidth}
  \includegraphics[width=8cm,clip]{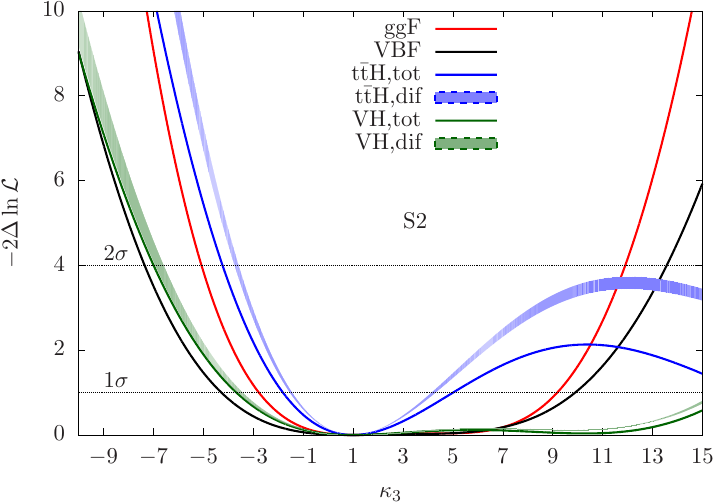}\caption{}\label{fig:s2-fit-k3-each}
  \end{subfigure}
  \caption{ $1\sigma$ and $2\sigma$ bounds on $\kt$ from single production processes, based on future projections for ATLAS-HL at 14 TeV. Left: only statistical uncertainty (S1). Right:  experimental systematic  uncertainty and theory uncertainty included (S2). 
 }
   \label{fig:fit-k3-individual}
\end{figure}
A first global fit on single Higgs channels has been performed in ref.~\cite{Degrassi:2016wml}  using  the 8 TeV LHC data, and a similar analysis has been applied to a future LHC scenario (CMS-HL-II) with $3000~{\rm fb}^{-1}$. Only total cross section
 information was used and especially the fit included  only $\lt$ as a variable. In ref.~\cite{DiVita:2017eyz} a first attempt to use differential rate information  provided in ref.~\cite{Degrassi:2016wml} was made by extrapolating the projections on total cross section from ATLAS-HL~\cite{ATL-PHYS-PUB-2013-014,ATL-PHYS-PUB-2014-011,ATL-PHYS-PUB-2014-012} with $3000~{\rm fb}^{-1}$.
 
Since no differential information is available in the measured data at the moment, we focus on the same future scenario at 14 TeV (ATLAS-HL) considered in ref.~\cite{DiVita:2017eyz}.
However, our results cannot be directly compared with those in ref.~\cite{DiVita:2017eyz}, since there are a few differences in the treatment of the inputs from experimental projections. Details are reported in Appendix~\ref{appic}, where we also carefully describe the procedure of the fit we performed and the assumptions on  the uncertainties.
In short, bounds on $\kt$, $\kappa_t$ and $\kappa_V$ are obtained by maximising a log-likelihood function.

We perform the fit considering two very different scenarios for the uncertainties. In the first scenario (S1), only the statistical uncertainty is included. This crude assumption corresponds to the ideal (and rather unrealistic) situation where theory and experimental systematic uncertainties are negligible. On the other hand, we exploit it for a direct comparison with the second scenario  (S2), where both theory and experimental systematic uncertainties are taken into account. At the differential level we performed the combination of the uncertainties via two different approaches that are described in detail in Appendix~\ref{appic}. For this reason differential results for this second scenario always appear as bands rather than lines, accounting the uncertainty related to the different assumptions on  the systematic and theory errors.

Before performing the global fit, we separately consider the different experimental inputs corresponding to $gg$F, VBF, $VH$ and $t \bar t H$ production\footnote{In this section when we refer to a production mode $X$ in fact we mean one of the different $X$-like categories in Tab.\ref{tab:in-ch}. As can be seen, in any $X$-like category the contribution of the actual $X$ process is in general dominant, so we can refer directly to it on the text for simplicity. Only the VBF-like category receives a non negligible contribution from $gg$F, which on the other hand has a $C_1$ very similar to VBF.} and we restrict to the configuration with $\kt$ only ($\kappa_t=\kappa_V=1$). We remind that different decay channels are entering for each of the production processes. Results are shown in Fig.~\ref{fig:fit-k3-individual}, where the plot on the left refers to  the scenario S1 and the plot on the right to  the scenario S2. For the case of $VH$ and $t \bar t H$ production dashed lines correspond to the fit of differential information; details on the binning are reported in Appendix~\ref{appic}.

 The different shapes of the curves for values smaller and larger than $\kt=1$ can be understood 
 from the behaviour of $\kt$ and $\kt^2$ terms in eqs.~\ref{ZHBSM} and \ref{predNLOBSM}.
 While for  $\kt<1$ both the $\kt$ and $\kt^2$ terms induce negative contributions in the production signal strengths, for $\kt>1$ there are large cancellations that suppress the effect of $\kt$.
 If we only include statistical uncertainty (S1) the $gg$F-like channel provides the best constraints for $\kt$ both for the region $\kt>1$ and $\kt<1$, where also $t \bar t H$ is giving strong constraints, which are not improved by the inclusion of differential information. A similar effect is visible also for $VH$; differential information does not lead to any significant improvement. 
On the other hand, in the region $\kt > 1$ we see a clear improvement due to differential information for $t \bar t H$, although bounds from this single production process are not sufficient to set a constraint in the region for $\kt>1$.
 
The plot on the right (S2) shows that including theory and experimental systematic uncertainties makes a difference. The $t \bar t H$ process is giving the strongest constraints in the region $\kt<1$ and receive improvements from the differential information, with a tiny dependence on the assumption made for the combination of the uncertainties. This difference is induced by the change of the $gg$F result moving from the scenario S1 to the scenario S2 rather than  by an improvement for $t \bar t H$. Note that, however, the impact of the differential information for $gg$F production is not known and while the exact calculation of the (two-)loop-induced effects from $\lt$ in $pp \to H j$ would be useful, it is currently out of reach.  Although constrains from $gg$F becomes much weaker in  the scenario S2, in the region $\kt>1$ they are still the strongest. At variance with $gg$F, $t \bar t H$ is in general very slightly affected by theory and systematic uncertainties since the dominant error is of statistical origin.  Regarding the bounds on $\kt$ from VBF-like and $VH$-like channels, they  are always worse than those from  $gg$F and $t \bar t H$, even when the differential information is used for $VH$. 
   \begin{figure}[t]
  \begin{subfigure}{0.5\linewidth}
   \includegraphics[width=8cm,clip]{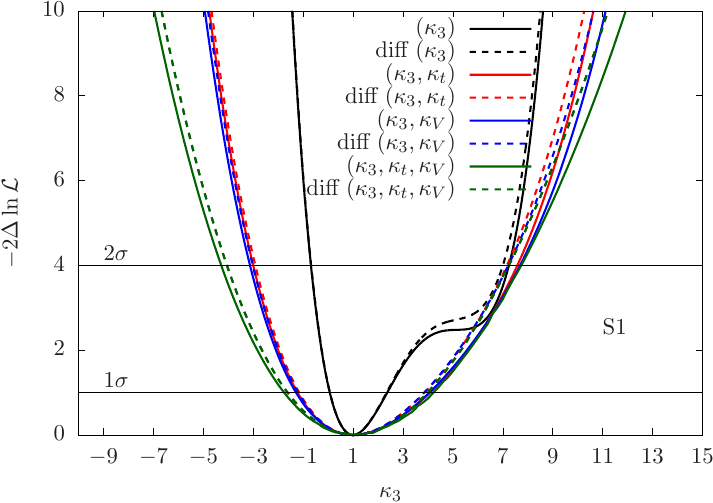}\caption{}
  \end{subfigure}
  \begin{subfigure}{0.5\linewidth}
   \includegraphics[width=8cm,clip]{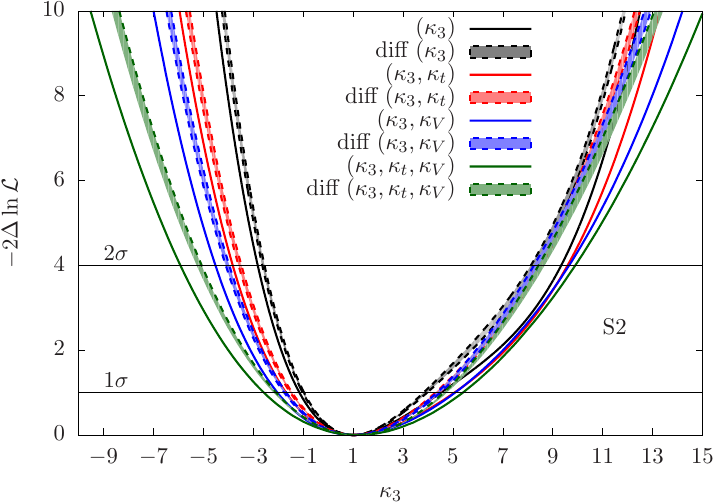}\caption{}
  \end{subfigure}
  \caption{
 $1\sigma$ and $2\sigma$ bounds on $\kt$ including all  production processes, based on future projections for ATLAS-HL at 14 TeV. Left: only statistical uncertainty (S1). Right:  experimental systematic  uncertainty and theory uncertainty included (S2).  
   }
   \label{fig:fit-k3}
\end{figure}
  \begin{figure}[!h]
\begin{subfigure}{0.5\linewidth}
  \includegraphics[width=8cm,clip]{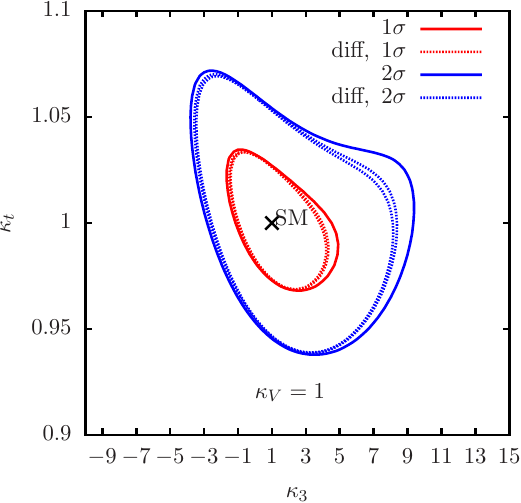}\caption{}
\end{subfigure} 
\begin{subfigure}{0.5\linewidth}
  \includegraphics[width=8cm,clip]{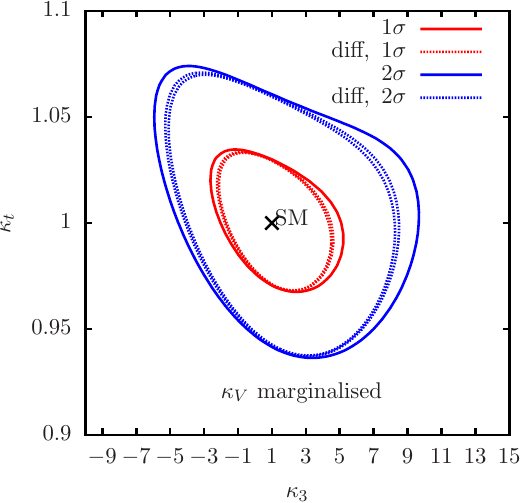}\caption{}
\end{subfigure} 
\begin{subfigure}{0.5\linewidth}
  \includegraphics[width=8cm,clip]{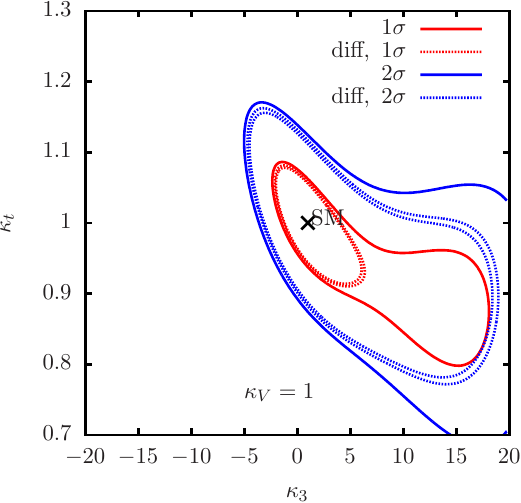}\caption{}
\end{subfigure} 
\begin{subfigure}{0.5\linewidth}
  \includegraphics[width=8cm,clip]{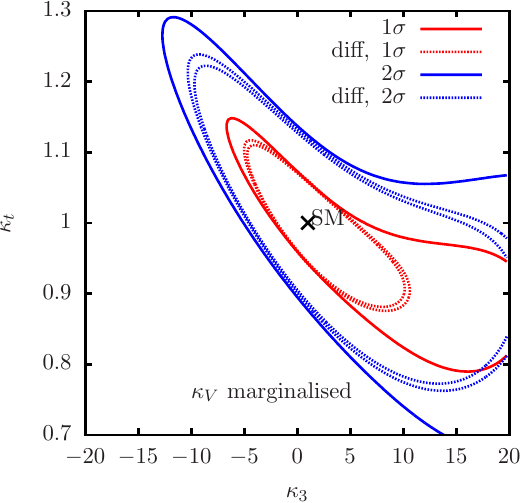}\caption{}
\end{subfigure} 
\caption{Bounds on $\kappa_t$ and $\kappa_3$. Left: $\kappa_V =1$. Right: $\kappa_V$ marginalised. Up: all channels considered in the  fit. Down: Only $VH$ and $t \bar t H$ considered in the fit.} 
\label{fig:fit-k3-kt-all}       
\end{figure}
 \begin{figure}[!h]
\begin{subfigure}{0.5\linewidth}
  \includegraphics[width=8cm,clip]{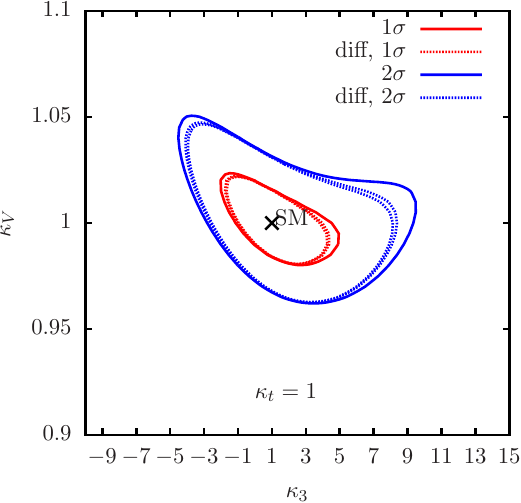}\caption{}
\end{subfigure} 
\begin{subfigure}{0.5\linewidth}
  \includegraphics[width=8cm,clip]{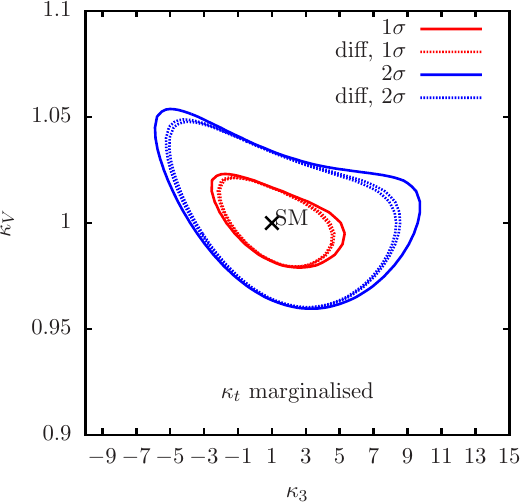}\caption{}
\end{subfigure} 
\begin{subfigure}{0.5\linewidth}
  \includegraphics[width=8cm,clip]{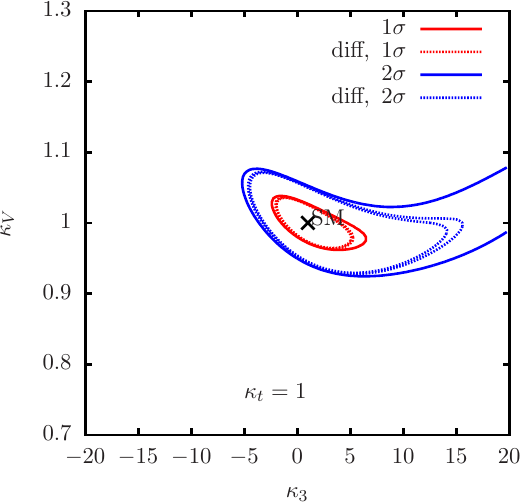}\caption{}
\end{subfigure} 
\begin{subfigure}{0.5\linewidth}
  \includegraphics[width=8cm,clip]{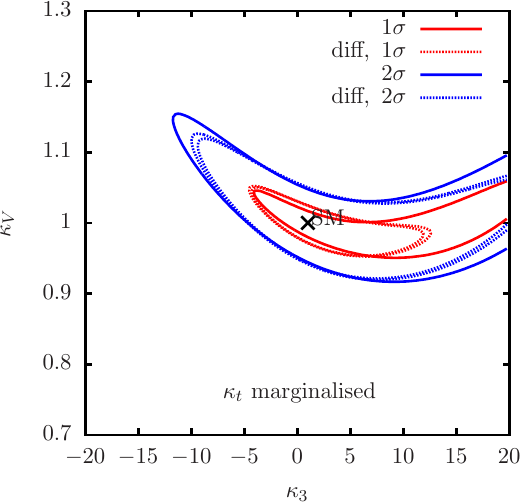}\caption{}
\end{subfigure} 
\caption{
Bounds on $\kappa_V$ and $\kappa_3$. Left: $\kappa_t =1$. Right: $\kappa_t$ marginalised. Up: all channels considered in the  fit. Down: Only $VH$ and $t \bar t H$ considered in the fit.} 
\label{fig:fit-k3-kv-all}       
\end{figure}

 Next, we perform the global fit including all the experimental data as input and taking into account the anomalous couplings $\kappa_t$ and $\kappa_V$.
 In Fig.~\ref{fig:fit-k3} we present bounds after combining all the production 
 channels, under different assumptions: i) only $\kt$ is anomalous, ii) $\kt,\kappa_t$ or $\kt,\kappa_V$ are anomalous, iii) all three parameters $\kt,\kappa_V,\kappa_t$ are anomalous.
In the presence of anomalous couplings other than $\kt$, we marginalise
 over them. The plot on the left refers to the scenario S1, only statistical uncertainties, and the one on the right to  the scenario S2, systematic and theory uncertainties included.
As we expect, in   the scenario S1 the differential information (dashed line) does not noticeably improve any of the constraints, while in  the scenario S2 in the region $\kt<1$ and especially in the region $\kt > 1$ differential information from $VH$ and $t \bar t H$ leads to a clear improvement of the constraints. What is instead not obvious, especially given the findings of ref.~\cite{DiVita:2017eyz}, is the effect induced by anomalous $\kappa_t$ and/or $\kappa_V$ terms to the fit. While constraints in the region $\kt < 1$ are relaxed, although not washed out completely, by the inclusion of one or two new degrees of freedom, in the region $\kt >1$ they are almost unaltered. In other words, in the scenario S2,  bounds in the region $\kt >1$ are more affected by differential information than by the addition of the $\kappa_t$ or $\kappa_V$ parameters.  Moreover, especially in the region $\kt <1$,  these two parameters alter the $\kt$ constraints more in the unrealistic scenario S1 than S2.
We describe in a bit of detail the observed features exploiting the information contained in Fig.~\ref{fig:fit-k3-individual}.

  In  the scenario S1 for $\kt<1$ the constraints are strongly affected by the inclusion of $\kappa_t$ and/or $\kappa_V$ since the global fit with only $\kt$ is completely dominated by $gg$F in that region. 
 For this process only the total cross section information is used in the fit, so that 
  a flat direction appears, {\it i.e.}, the fit is dominated by one input~\footnote{Note  we have three decay channels
 for $gg$F that are almost fully controlled by $k_V$, namely $WW^*, VV$ and  $\gamma \gamma$. Indeed, also for $H\to \gamma\gamma$ 
 the contribution from top-quark loop is known to be much smaller than $W$-loop contribution.}, which is sufficient for setting constraints on only $\kt$ but not at the same time on $\kt$ and $\kappa_t,\kappa_V$. 
 To resolve this degeneracy more constraining information must be added to the fit. Indeed, the constraints with two parameters  ($\kt,\kappa_t$ or $\kt,\kappa_V$) or three $(\kt, \kappa_t,\kappa_V)$ are in the region of the constraints from VBF and $t \bar t H $ in Fig.~\ref{fig:fit-k3-individual}.
  
 The previous argument cannot be applied to the region $\kt>1$ for  the scenario S1, where the bounds in the global fit with only $\kt$ are not completely dominated by $gg$F. Indeed the $t \bar t H$ (and in a smaller way the VBF) contribution is not negligible in that region, as can be seen from  the left plot of Fig.~\ref{fig:fit-k3-individual}. Moreover, at variance with $gg$F production, there is not a large background  in $t \bar t H$  production for the experimental signatures involving the Higgs to $\mu^+ \mu^-$ decay, whose branching ratio has a different $\kappa_V$ and $\kappa_t$ dependence w.r.t. $\gamma \gamma $ and $V V^*$ decays, and for values $\kt\sim 8$ the impact of decays is more relevant. For this reason $t \bar t H$ and $gg$F are sufficient for constraining one, two or three parameters, with negligible difference when parameters other than $\kt$ are marginalised. We explicitly verified this feature.  
 
Moving to the scenario S2, the plot on the right where all uncertainties are included, for $\kt<1$ the bounds are dominated by $t \bar t H$ channel. For this reason there is  a smaller dependence on the number of parameters considered in the fit and a larger sensitivity to the differential information, which is present for the same reason also in the region $\kt > 1$.   

It is clear that the role of the $gg$F is essential when the impact of differential information is investigated in the global fit. When $gg$F is dominant, since there is no differential dependence, it masks the relevance of differential distributions. On the other hand, when $t \bar t H $ is dominant, the differential information becomes relevant. Above all, one should bear in mind  that the impact of $\kt$ on $gg$F distributions has not been calculated because of technical reasons; the exact two-loop calculation is beyond the current technology, but could be relevant too. To this purpose, in the following we look at constraints in the $(\kt, \kappa_t)$ and $(\kt, \kappa_V)$ plane with and without the contributions from VBF and $gg$F, which hides the impact of the differential information. We consider only the scenario S2, which is more realistic.

 In Fig.~\ref{fig:fit-k3-kt-all}, we provide $1\sigma$ and $2\sigma$ contours in $(\kt,\kappa_t)$ plane
 without (left) and with (right) anomalous $\kappa_V$, which is anyway marginalised. Upper plots includes all the production channels, whereas in the lower ones only $VH$ and $t \bar t H$  enter. Analogous plots are provided in Fig.~\ref{fig:fit-k3-kv-all} for the $(\kt,\kappa_V)$, without and with  anomalous $\kappa_t$. 
First of all, one can note that due to the $\kappa_t$ dependence of the gluon fusion channel and $t \bar t H$ channel, in the upper plots the constraints on $\kt$ in presence 
of $\kappa_t$  (Fig.~\ref{fig:fit-k3-kt-all}) are stronger than those in presence of $\kappa_V$ (Fig.~\ref{fig:fit-k3-kv-all}).~\footnote{ For the same reason, comparing these results with those that would be obtained in the scenario S1, one may also find that after including all uncertainties, the bounds on $\kappa_t$ are enlarged more significantly than those on $\kappa_V$,
 since the dominant contribution to bounds on $\kappa_t$ is $gg$F and $t \bar t H$,
 and the experimental systematic  uncertainty and theory uncertainty
 are much larger than statistical uncertainty for $gg$F.} Also, in the upper plots, having two independent parameters (left) or marginalising on an additional third one (right) does not lead to qualitatively significant differences. As discussed also before, the impact of the differential information is more important.
 
 If we consider the lower plots the situation is very different. First, constraints with two or three parameters are qualitatively different. Second, the impact of the distributions is much more relevant. In the lower-left plot of Fig.~\ref{fig:fit-k3-kv-all} a flat direction is clearly resolved by differential information. The bottom-line is that by changing the number of free parameters and the number of inputs entering in the fit, the relevance of differential distributions and the sensitivity of the $\kt$-limits on additional parameters can be considerably altered. The range of the lower plots is much larger than in the upper plots, for this reason the exclusion of $\kt^2 \kappa_t^2$ and/or $\kt^2 \kappa_V^2$ terms from eq.~\eqref{mui} would lead to visible effects to the 2$\sigma$ contours, anyway without altering the qualitative information.  
 \section{Conclusion}\label{sec:conclusion}
 We have studied one-loop $\lt$ effects  for all the relevant single Higgs production modes at the LHC ($gg{\rm F}$, ${\rm VBF}$, $VH$, $t\bar tH$, $tHj$) and decays ($\gamma\gamma$, $VV^*$, $ff$, $gg$), extending and completing the results presented in ref.~\cite{Degrassi:2016wml}.  In particular, we have calculated differential results for VBF, $VH$, $t \bar t H$ and $tHj$  production and $H\to 4\ell$ decay. We have developed an automated code, which has been made public, for generating events including one-loop $\lt$ effects. All the distributions that may be potentially affected by anomalous values of $\lt$  have been scrutinised: differential level results for $t \bar t H$ production, $H\to 4\ell$ decay and also for the $tHj$ process have been presented here for the first time. 
 
 We find that the production modes with a large kinematic dependence on $\lt$  are $VH$, $t \bar t H$ and $tHj$. In particular, $VH$ and $t \bar t H$ can  provide  additional sensitivity on $\lt $ at differential level. For these two channels we have consistently combined complete SM NLO EW corrections with anomalous $\lt$-induced effects at differential level. The same combination has been performed, at inclusive level, also for all the other production processes.  We have verified the robustness of our strategy: NLO EW corrections are essential for a precise determination of anomalous $\lt$ effects, but  they do not jeopardise the efficiency of indirect $\lt$ determination. We note that NLO EW corrections to $tHj$ in the SM were unknown and have been calculated for the first time too.

Finally, we have performed a fit for $\kt$ based on the future projections of ATLAS-HL for single-Higgs production and decay at 14 TeV~\cite{ATL-PHYS-PUB-2013-014, ATL-PHYS-PUB-2014-012}. We have considered the effects induced on the fit by additional degrees of freedom, in particular,  anomalous Higgs couplings with the vector bosons and/or the top quark.  We have found  that in a global fit, including all the possible production and decay channels, two additional degrees of freedom such as those considered here do not preclude the possibility of setting sensible $\lt$ bounds, especially, they have a tiny impact on the upper bound for positive $\lt$ values. On the contrary, the role of differential information may be relevant, critically depending on the assumptions on the future experimental and theoretical uncertainties. We have also shown that the relevance of differential distributions and the sensitivity on $\kt$ can be considerably altered by varying  the  relation among the number of free parameters and the number of inputs entering in the fit. 
  
Our results clearly illustrate the complementarity of precise single-Higgs measurements and double Higgs searches at the LHC for constraining $\lt$ with the current and future accumulated luminosity.  We therefore encourage experimental collaborations to use the MC tool provided here for performing $\lt$ determination via single Higgs measurements, taking into account all the possible correlations among theory and experimental uncertainties of the different production and decay channels.

 \section*{Acknowledgements}
We are grateful to the LHCHXSWG for always providing motivation and a stimulating environment.
We acknowledge many enlightening discussions and continuous collaboration on the Higgs self coupling determination with Giuseppe Degrassi, Pier Paolo Giardino, Stefano Di Vita and Christophe Grojean.  This work is supported in part by the ``Fundamental interactions" convention FNRS-IISN 4.4517.08. The work of D.P. is supported by the Alexander von Humboldt Foundation, in the framework of the Sofja Kovalevskaja Award Project ``Event Simulation for the Large Hadron Collider at High Precision''. The work of A.S. is supported by the MOVE-IN Louvain Cofund grant. The work of X.Z. is supported by the European Union Marie Curie Innovative Training Network MCnetITN3 722104.   
 \appendix

 \section{Details about the fit and the data input}\label{appic}
We describe here in detail how we performed the fit discussed in section~\ref{sec:global-fit},
 and how the input from  refs.~\cite{ATL-PHYS-PUB-2013-014, ATL-PHYS-PUB-2014-012} is treated. In Tab.~\ref{tab:in-ch} we report  numbers derived from these references. The notation``$X$-like'', which is present also in the original reference, means that the Higgs boson is likely to be produced through the $X$ production mechanism,
 after applying appropriate cuts.
 For example, VBF-like means that in this channel
 Higgs bosons are likely to be produced through vector boson fusion.
 However, there are other production mechanism that will contribute to this channel;
  in VBF-like channels the contribution from $gg$F are comparable to VBF.
 At variance with ref.~\cite{DiVita:2017eyz}, we take into account that a given $X$-like production channel  can receive contributions from all the four $Y$ production mechanisms. Moreover, we did not include in our analysis results from ref.~\cite{ATL-PHYS-PUB-2014-011}, where $WH$ and $ZH$ production with $H\to b\bar b$ decays have been considered. We have anyway verified that their impact is negligible, due to the large contribution from the background. 
  
  In order to correctly taking into account that different production mechanisms contribute to a given production-like channel, we calculate the number of events in each channel as
\begin{equation}
    N_{l,f}=N_{l, f}^{\rm bkg}+\sum_{i}\mu_i^f N_{l,i,f}^{\rm SM}\, ,\label{eq:tot-evt}
\end{equation}
where $l\in\{$$gg$F-like, VBF-like, $WH$-like, $ZH$-like, $t \bar t H$-like$\}$ is the
production-like channel, $f$ is the decay channel and $i$ is the actual production mechanism.
Thus, for a given production-like channel $l$ with decay $f$ the total number of events is given by  the number  of background events $N_{l, f}^{\rm bkg}$ and 
the sum of all the production mechanisms $i$ contributing to $l$, {\it i.e.}, the number of SM events $N_{l,i,f}^{\rm SM}$ multiplied by the corresponding  signal strength $\mu_i^f$ defined in Eq.~\eqref{signalstre}.
The value for $N_{l,i,f}^{\rm SM}$ and $N_{l,f,{\rm bkg}}$ are listed in Tab.~\ref{tab:in-ch}. The symbol $1\ell$($2\ell$) means one(two) lepton(s) observed in the final state,
 ``lept.''(``semi-lept.'') means leptonic(semi-leptonic) decay of $\tau^+\tau^-$ pair, and 0j(1j) means 0(1) extra jet.

In order to perform the fit, we adopt Gaussian distribution for the events as approximation, and define likelihood function as following:
\begin{align}
    \mathcal{L}=\prod_{l,f}\frac{1}{\sqrt{2\pi\sigma^2_{l,f}}}\exp\left[-{\frac{\left(N_{l,f}-N_{l,f}^{{\rm SM}}\right)^2}{2\sigma^2_{l,f}}}\right]\, . \label{likelihood}
\end{align}
In eq.~\eqref{likelihood} the quantity $N_{l,f}^{{\rm SM}}$ is simply $N_{l,f}$ where all the signal strengths have been set to one, while  
$\sigma_{l,f}$ is the total (absolute) uncertainty for each channel obtained by summing in quadrature  statistical ($\sigma_{l,f}^{ \rm stat}$), theory ($\sigma_{l,f}^{  \rm th}$) and experimental systematic $(\sigma_{l,f}^{ \rm sys})$  uncertainties. 
The statistical and theory uncertainty are calculated as
\begin{align}
    (\sigma_{l,f}^{\rm stat})^2=N_{l,f}^{{\rm SM}}\, ,\\
    (\sigma_{l,f}^{\rm th})^2=\sum_{i} (N_{l,i,f}^{{\rm SM}} \epsilon_{{\rm th},i})^2\, , 
\end{align}
where $\epsilon_{i }^{\rm th}$
is the relative theory uncertainty for each production mechanism $i$
and we treat it  as uncorrelated with the other different
production mechanism.
We list the theory uncertainty in Tab.~\ref{tab:th-err}, they are taken from the YR4~\cite{deFlorian:2016spz}.
Concerning experimental systematic  uncertainties, we list them directly in Tab.~\ref{tab:in-ch},  based on estimation from \cite{ATL-PHYS-PUB-2013-014, ATL-PHYS-PUB-2014-012} and expressed directly as number of events and not as relative numbers. In the scenario S1 discussed in sec.~\ref{secresults} we set $\sigma_{l,f}^{ \rm th}=\sigma_{l,f}^{ \rm sys}=0$ while in the scenario S2 we keep these uncertainties.

 \begin{center}
 \renewcommand{\arraystretch}{1.1}
 \begin{table}[t]
     \small
     \begin{tabular}{|c|c|c|c|c|c|c|c|c|}
         \hline
         \multicolumn{2}{|c|}{Category} & ggF & VBF & $WH$ & $ZH$ & $t \bar t H$ & Backgrounds & sys.\\
         \hline
         \multirow{5}{*}{$ ZZ^{*}$}
         & $gg$F-like & 3380 & 274 & 77 & 53 & 25 & 2110 & 283\\
         & VBF-like & 41 & 54 & 0.7 & 0.4 & 1.0 & 4.2 & 7.4\\
         & $WH$-like & 22 & 6.6 & 25 & 4.4 & 8.8 & 1.3 & 4.9\\
         & $ZH$-like & 0.0 & 0.0 & 0.01 & 4.4 & 1.3 & 0.06 & 0.41\\
         & $t \bar t H$-like & 3.1 & 0.6 & 0.6 & 1.1 & 30 & 1.6 & 3.2\\
         \hline
         \multirow{6}{*}{$ \gamma\gamma$}
         & $gg$F-like & \num{7.51e4} & \num{5.66e3} & 0 & 0 & 0 & \num{4.06e6} & \num{2.5e3}\\
         & VBF-like & 63.9 & 149 & 0 & 0 & 0 & 802 & 6.5\\
         & $WH$-like & 15.9 & 9.08 & 163 & 2.27 & 15.9 & 995 & 7.4\\
         & $ZH$-like & 0 & 0 & 0 & 23.0 & 3.13 & 22.8 & 0.85\\
         & $t \bar t H$-like,~$1\ell$ & 6.75 & 0 & 11.3 & 4.5 & 200 & 428 & 6.9\\
         & $t \bar t H$-like,~$2\ell$ & 0 & 0 & 0 & 0.38 & 18.5 & 48.3 & 0.98\\
         \hline
         \multirow{3}{*}{$WW^{*}$}
         & $gg$F-like,~0j & 40850 & 990 & 0 & 0 & 0 & 366450 & \num{9.5e3}\\
         & $gg$F-like,~1j & 20050 & 2325 & 0 & 0 & 0 & 259610 &\num{1.1e4}\\
         & VBF-like & 90 & 500 & 0 & 0 & 0 & 1825 & \num{1.6e2}\\
         \hline
         \multirow{2}{*}{$ \tau^{+}\tau^{-}$}
         & VBF-like,~lept. & 0 & 147 & 0 & 0 & 0 & 190 & 10\\
         & VBF-like,~semi-lept. & 0 & 297 & 0 & 0 & 0 & 1610 & 21 \\
         \hline
         \multirow{2}{*}{$ \mu^{+}\mu^{-}$}
         & $gg$F-like & \num{1.51e4} & \num{1.25e3} & 450 & 270 & 180 & \num{5.64e6} & 630\\
         & $t \bar t H$-like & 0 & 0 & 0 & 0 & 33 & 22 & 1.7\\
         \hline
     \end{tabular}
 \caption{ Number of signal and background events in each $X$-like production channel and decay for 
          ATLAS-HL at 14 TeV LHC with $3000~{\rm fb^{-1}}$ luminosity. In the first four column there are the numbers corresponding to $N_{l,i,f}^{\rm SM}$, $l$ and $f$ can be read on the left, $i$ is at the top. $N_{l,f}^{ \rm bkg}$ and $\sigma_{l,f}^{ \rm sys}$ are displayed in the fifth and sixth columns, respectively. 
 \label{tab:in-ch}}
 \end{table}
\end{center}

So far we discussed the case of total cross section; numbers listed in Tab.~\ref{tab:in-ch} are for inclusive Higgs production.
In the case of differential distributions for $VH$ and $t \bar t H$, eq.~\eqref{likelihood} can be generalised by independently considering each bin for these two processes. In practice, we split $N_{l,f}$ into several  $p_T(H)$ bins and for each bin $j$ the number of events is given by 
\begin{equation}
    N_{l,f,j}=N_{l, f,j}^{\rm bkg}+\sum_{i}\mu_{i,j}^f N_{l,i,f,j}^{\rm SM}\, .\label{eq:diff-evt}
\end{equation}

In eq.~\eqref{eq:diff-evt} we made the following assumptions:
\begin{align}
N_{t \bar t H{\rm -like},i,f,j}^{{\rm SM}}= r_{j}^{t \bar t H} N_{t \bar t H{\rm -like},i,f}^{{\rm SM}}\, ,\\
N_{ZH{\rm -like},i,f,j}^{{\rm SM}}= r_{j}^{ZH} N_{ZH{\rm -like},i,f}^{{\rm SM}} \, , \\
N_{WH{\rm -like},i,f,j}^{{\rm SM}}= r_{j}^{WH} N_{WH{\rm -like},i,f}^{{\rm SM}} \, ,
\end{align}
where $r_{j}^{i}(i=t \bar t H,ZH,WH)$ is the ratio of the cross section of the bin $j$ with  the total cross section for process $i$. In other words,  for each production-like mode we use $r_{j}^{i}$ only form the dominant production process and decay, {\it i.e.}, $N_{X{\rm -like},i,f,j}\to r_{j}^{X} $. The same assumption is made for the background. 
 
NLO EW $K-$factors at  differential level are considered and used for computing the $\mu_{i,j}^f$, which is simply the signal-strength  prediction for each bin $j$.  The chosen binning and the corresponding $r_{j}^{i}$, $K^{\rm EW}$ and $C_1$ for each bin can be found in Tab.~\ref{tab:ratio-xs}. For each bin, the statistical uncertainty  is determined via its number of events and the relative theory uncertainty is assumed to be the same at the inclusive level.
We may overestimate or underestimate the theory uncertainty, since correlations are certainly present in the different bins but also in the different processes.

Concerning the experimental systematic uncertainty, we consider two cases. Either we scale it as $r_{j}^{i}$,  so that in each bin is preserved the relative uncertainty that is present at the level of the total cross section, or
as $\sqrt{r_j^{i}}$, so that the sum in quadrature of the uncertainties for each bin is giving the value of the uncertainty for the total cross section, as in the case of the statistical uncertainty. The difference between the two approaches is small, as can be seen in sec.~\ref{sec:global-fit}; for the plots concerning the S2 scenario, the two approaches correspond to the border of the bands.
\begin{table}
\renewcommand{\arraystretch}{1.1}
\begin{center}
    \begin{tabular}{|c|c|c|c|c|c|c|}
        \hline
        production & $gg$F & VBF & $WH$ & $ZH$ & $t  \bar t H$ \\
        \hline
        $\epsilon_{th}$(\%) & 5 & 2 & 2 & 4 & 8\\
        \hline
    \end{tabular}
    \caption{Theory uncertainty for different production channels.\label{tab:th-err}}
\end{center}
\end{table}
\begin{table}
\renewcommand{\arraystretch}{1.3}
    \begin{center}
        \begin{tabular}{|c|c|c|c|c|c|}
    \hline
    $p_T(H)(\textrm{GeV})$ & 0-50 & 50-100 & 100-150 & 150-200 & 200+ \\
    \hline
    $r^{t\bar{t}H}$ & 0.173 & 0.312 & 0.227 & 0.128 & 0.159 \\
    \hline
    $r^{ZH}$ &0.336 &0.379 &0.165 &0.065 &0.055 \\
    \hline
    $r^{WH}$ &0.345 &0.375 &0.162 &0.064 &0.055 \\
    \hline
     \hline
    $K^{\rm EW}_{\rm SM}(t\bar{t}H$) & 1.041 & 1.027 & 1.012 & 0.999 & 0.925 \\
    \hline
    $K^{\rm EW}_{\rm SM}$($ZH$) &0.950 &0.946 &0.952 &0.939 &0.889 \\
    \hline
    $K^{\rm EW}_{\rm SM}$($WH$) &0.941 &0.935 &0.920 &0.905 &0.840 \\
    \hline
     \hline
    $C_1(t\bar{t}H)(\%)$ & 5.14 & 4.38 & 3.33 & 2.41 & 1.21 \\
    \hline
    $C_1(ZH)(\%)$ & 1.83 & 1.21 & 0.61 & 0.25 & 0.001 \\
    \hline
    $C_1(WH)(\%)$ & 1.56 & 1.04 & 0.53 & 0.22 & 0.007 \\
    \hline
        \end{tabular}
        \caption{Ratio of the cross section, NLO EW K-factor, and $C_1$ in different bins of $p_T(H)$  for $t \bar t H$, $ZH$ and $WH$.\label{tab:ratio-xs}}
    \end{center}
\end{table}

 \bibliographystyle{ieeetr}
 \bibliography{paper_lambda_diff} 
 
\end{document}